\documentclass[11pt]{article}
\usepackage{bbm}
\usepackage{graphicx}
\usepackage{amsmath}
\usepackage{amssymb}
\usepackage{caption2}
\begin{document}
\bibliographystyle{prsty}
\begin{center}
{\large {\bf \sc{ Analysis of the vertexes $\Xi_Q^*\Xi'_Q V$,
 $\Sigma_Q^*\Sigma_Q V$ and radiative decays
$\Xi_Q^*\to \Xi'_Q \gamma$, $\Sigma_Q^*\to \Sigma_Q \gamma$ }}} \\[2mm]
Zhi-Gang Wang \footnote{E-mail,wangzgyiti@yahoo.com.cn. }    \\
 Department of Physics, North China Electric Power University, Baoding 071003, P. R. China \\
\end{center}

\begin{abstract}
In this article, we study the vertexes  $\Xi_Q^*\Xi'_Q V$ and
 $\Sigma_Q^*\Sigma_Q V$ with the light-cone QCD sum rules, then assume  the vector meson
dominance of the intermediate $\phi(1020)$, $\rho(770)$ and
$\omega(782)$, and calculate the radiative decays $\Xi_Q^*\to \Xi'_Q
\gamma$ and  $\Sigma_Q^*\to \Sigma_Q \gamma$.

\end{abstract}


\section{Introduction}

The charm and bottom baryons which  contain a heavy quark and two
light quarks are particularly interesting for studying dynamics of
the light quarks in the presence  of a heavy quark. They  serve as
an excellent ground for testing predictions of the quark models and
heavy quark symmetry \cite{ReviewH1,ReviewH2}. The three light
quarks form an $SU(3)$ flavor triplet ${\bf 3}$,  two light quarks
can form diquarks of a symmetric sextet  and an antisymmetric
antitriplet, i.e. ${\bf 3}\times {\bf 3}={\bf \bar 3}+{\bf 6}$. For
the $S$-wave charm baryons, the ${1\over 2}^+$ antitriplet states
($\Lambda_c^+$, $\Xi_c^+,\Xi_c^0)$,  and the ${1\over 2}^+$ and
${3\over 2}^+$ sextet states ($\Omega_c,\Sigma_c,\Xi'_c$) and
($\Omega_c^*,\Sigma_c^*,\Xi^*_c$) have been well established; while
the corresponding bottom  baryons are far from complete, only the
$\Lambda_b$, $\Sigma_b$, $\Sigma_b^*$, $\Xi_b$, $\Omega_b$ have been
observed \cite{PDG}. Furthermore, several new excited charm baryon
states have been observed by the BaBar, Belle and CLEO
Collaborations, such as $\Lambda_c(2765)^+$, $\Lambda_c^+(2880)$,
$\Lambda_c^+(2940)$, $\Sigma_c^+(2800)$, $\Xi_c^+(2980)$,
$\Xi_c^+(3077)$,
  $\Xi_c^0(2980)$ , $\Xi_c^0(3077)$ \cite{ShortRV1,ShortRV2,ShortRV3}.

In Ref.\cite{WangVMD}, we assume the charm  mesons $D_{s0}(2317)$
and  $D_{s1}(2460)$ with the spin-parity $0^+$ and $1^+$
respectively are
  the conventional $c\bar{s}$
states, and  calculate the strong coupling constants $\langle D_s^*
\phi | D_{s0} \rangle$ and$\langle D_s \phi | D_{s1} \rangle$  with
the light-cone QCD sum rules, then take the vector meson dominance
of the intermediate $\phi(1020)$,  study the radiative decays
$D_{s0}\to D_s^* \gamma $ and $D_{s1}\to D_s \gamma $. In
Refs.\cite{Wang0809Omega,Wang0704}, we   calculate the masses and
the pole residues of the $\frac{1}{2}^+$ heavy baryons $\Omega_Q$
and the $\frac{3}{2}^+$ heavy baryons $\Omega_Q^*$   with the QCD
sum rules.  Moreover,  we study the vertexes $\Omega_Q^*\Omega_Q
\phi$ with the light-cone QCD sum rules, then assume  the vector
meson dominance of the intermediate $\phi(1020)$, and calculate the
radiative decays $\Omega_Q^*\to \Omega_Q \gamma$ \cite{Wang0909}.

 In this article, we
extend our previous works to  study the vertexes $\Xi_Q^*\Xi'_Q V$
and  $\Sigma_Q^*\Sigma_Q V$ with the light-cone QCD sum rules
\footnote{The results of the strong coupling constants among the
nonet vector mesons, the octet baryons and  the decuplet baryons
will be presented  elsewhere.}, then assume  the vector meson
dominance of the intermediate $\phi(1020)$, $\rho(770)$ and
$\omega(782)$, and calculate the radiative decays $\Xi_Q^*\to\Xi'_Q
\gamma$ and
 $\Sigma_Q^*\to\Sigma_Q \gamma$ to complete our works on radiative decays among the  ${1\over 2}^+$ and ${3\over 2}^+$
sextet states ($\Omega_Q,\Sigma_Q,\Xi'_Q$) and
($\Omega_Q^*,\Sigma_Q^*,\Xi^*_Q$).
  In Ref.\cite{Aliev0901}, Aliev et al study  the radiative  decays
$\Sigma_Q^*\to\Sigma_Q\gamma$, $\Xi_Q^*\to\Xi_Q\gamma$ and
$\Sigma_Q^*\to\Lambda_Q\gamma$
 with the light-cone
QCD sum rules, where the light-cone distribution amplitudes of the
photon are used.

The light-cone QCD sum rules  carry out the operator product
expansion near the light-cone $x^2\approx 0$ instead of the short
distance $x\approx 0$, while the nonperturbative hadronic matrix
elements  are parameterized by the light-cone distribution
amplitudes   instead of  the vacuum condensates
\cite{LCSR89,LCSR,LCSRreview}. The nonperturbative
 parameters in the light-cone distribution amplitudes are calculated with  the conventional QCD  sum rules
 and the  values are universal. Based on the quark-hadron duality, we
can obtain copious information about the hadronic parameters at the
phenomenological side \cite{LCSRreview,SVZ79,PRT85}.   The  $\rho
NN$, $\rho\Sigma\Sigma$, $\rho\Xi\Xi$ and other strong coupling
constants of the nonet vector mesons with the octet baryons have
been calculated using the light-cone QCD sum rules
\cite{Zhu99,Wang0701,Aliev0905}. In Refs.\cite{Aliev2006,Aliev0908},
Aliev et al study the strong coupling constants of the pseudoscalar
octet mesons with the octet (decuplet) baryons   comprehensively. In
Refs.\cite{Wang0707,Wang0809}, we study the strong decays
$\Delta^{++} \to p \pi$, $\Sigma^*\to \Sigma \pi$ and $\Sigma^*\to
\Lambda \pi$ using the light-cone QCD sum rules. Moreover, the
coupling constants of the vector mesons $\rho$ and $\omega$ with the
baryons are studied with the external field QCD sum rules
\cite{Erkol2006}. Recently, the strong coupling constants among the
light vector mesons and the heavy baryons are calculated with the
light-cone QCD sum rule in the leading order of heavy quark
effective theory \cite{Zhu0909}.

The article is arranged as follows:  we derive the strong coupling
constants  $g_1$, $g_2$ and $g_3$ of the vertexes $B_Q^*B_Q V$ with
the light-cone QCD sum rules in Sect.2; in Sect.3, we present the
 numerical results and discussions; and Sect.4 is reserved for our
conclusions.
\section{ The vertexes  $B_Q^*B_Q V$  with light-cone QCD sum rules}
We parameterize the vertexes $B_Q^*B_Q \phi$, $B_Q^*B_Q \rho$ and
$B_Q^*B_Q \omega$ with three tensor structures  due to  Lorentz
invariance  and introduce three strong  coupling constants $g_1$,
$g_2$ and $g_3$ \cite{ANN1973},
\begin{eqnarray}
\langle B_Q(p+q)|B_Q^*(p)\phi(q) \rangle&=&\overline{U}(p+q) \left[
g_1(q_\mu\not\!\!{\epsilon}-\epsilon_\mu \not\!\!{q})\gamma_5+g_2
(P\cdot\epsilon q_\mu-P\cdot q \epsilon_\mu) \gamma_5
\right.\nonumber\\
&&\left. +g_3(q\cdot \epsilon q_\mu-q^2
\epsilon_\mu)\gamma_5\right]U^\mu(p) \, ,\\
\langle B_Q(p+q)|B_Q^*(p)\rho_0/\omega(q)
\rangle&=&\frac{1}{\sqrt{2}}\overline{U}(p+q) \left[
g_1(q_\mu\not\!\!{\epsilon}-\epsilon_\mu \not\!\!{q})\gamma_5+g_2
(P\cdot\epsilon q_\mu-P\cdot q \epsilon_\mu) \gamma_5
\right.\nonumber\\
&&\left. +g_3(q\cdot \epsilon q_\mu-q^2
\epsilon_\mu)\gamma_5\right]U^\mu(p)  \, ,
\end{eqnarray}
where the $U(p)$ and $U_\mu(p)$ are the Dirac spinors of the heavy
baryon states $B_Q$ ($\Xi'_Q$, $\Sigma_Q$) and $B^*_Q$ ($\Xi_Q^*$,
$\Sigma_Q^*$) respectively,   the $\epsilon_\mu$ is the polarization
vector of the mesons $\phi(1020)$, $\rho(770)$ and $\omega(782)$,
and $P=\frac{2p+q}{2}$.

In the following, we write down the
 two-point correlation functions  $\Pi_\mu(p,q)$,
\begin{eqnarray}
\Pi_\mu^{\phi/\rho_0}(p,q)&=&i \int d^4x \, e^{-i p \cdot x} \,
\langle 0 |T\left\{ J(0)\bar{J}_\mu(x)\right\}|\phi/\rho_0(q)\rangle \, , \\
J^\Xi(x)&=& \epsilon^{ijk}  q^T_i(x)C\gamma_\mu s_j(x) \gamma_5 \gamma^\mu Q_k(x)  \, ,  \nonumber \\
J^\Sigma(x)&=& \epsilon^{ijk}  q^T_i(x)C\gamma_\mu q'_j(x) \gamma_5 \gamma^\mu Q_k(x)  \, ,  \nonumber \\
J^\Xi_\mu(x)&=& \epsilon^{ijk}  q^T_i(x)C\gamma_\mu s_j(x)  Q_k(x)\, ,  \nonumber \\
J^\Sigma_\mu(x)&=& \epsilon^{ijk}  q^T_i(x)C\gamma_\mu q'_j(x)
Q_k(x)\, ,
\end{eqnarray}
where $Q=c,b$ and $q,q'=u,d$, the $i,j,k$ are color indexes,  the
Ioffe type  heavy baryon currents $J(x)$ ($J^{\Xi}(x)$,
$J^\Sigma(x)$) and $J_\mu(x)$ ($J^\Xi_\mu(x)$, $J^\Sigma_\mu(x)$)
 interpolate the $\frac{1}{2}^+$ baryon states $\Xi'_Q$, $\Sigma_Q$ and the  $\frac{3}{2}^+$
 baryon states  $\Xi_Q^*$, $\Sigma_Q^*$,  respectively, the external  vector states $\phi(1020)$
 and $\rho(770)$ have the
four momentum $q_\mu$ with $q^2=M_{\phi/\rho}^2$. The quark
constituents of the vector mesons $\rho_0$ and $\omega$ are
$\frac{1}{\sqrt{2}}\left(|\bar{u}u\rangle-|\bar{d}d\rangle\right)$
and
$\frac{1}{\sqrt{2}}\left(|\bar{u}u\rangle+|\bar{d}d\rangle\right)$
respectively, the isospin triplet meson $\rho_0$ and isospin singlet
meson $\omega$ have approximately   degenerate masses. We assume
that the vector mesons $\rho_0$ and $\omega$ have similar light-cone
distribution amplitudes, and obtain the corresponding strong
coupling constants
 by symmetry considerations, as
the $\omega$-meson light-cone distribution amplitudes have not been
explored yet, see Appendix A for detailed discussions.

Basing on the quark-hadron duality \cite{SVZ79,PRT85}, we can insert
a complete set  of intermediate hadronic states with the same
quantum numbers as the current operators $J(x)$ and $J_\mu(x)$ into
the correlation functions $\Pi_{\mu}(p,q)$  to obtain the hadronic
representation. After isolating the ground state contributions from
the pole terms of the heavy  baryons $\Xi'_Q$, $\Sigma_Q$ and
$\Xi_Q^*$, $\Sigma_Q^*$, we get the following results,
\begin{eqnarray}
\Pi_{\mu}^{\phi/\rho_0}(p,q)&=&\frac{\langle0| J(0)|
B_Q(q+p)\rangle\langle B_Q(q+p)| B_Q^*(p) \phi/\rho(q) \rangle
\langle B_Q^*(p)|\bar{J}_\mu(0)| 0\rangle}
  {\left[M_{B_Q}^2-(q+p)^2\right]\left[M_{B_Q^*}^2-p^2\right]}  + \cdots \nonumber \\
&=& \frac{\lambda_{B_Q}\lambda_{B_Q^*}}
{\left[M_{B_Q}^2-(q+p)^2\right]\left[M_{B_Q^*}^2-p^2\right]}
\left\{g_1 \left[M_{B_Q}+M_{B_Q^*}
\right]\!\not\!{\epsilon}\!\not\!{p} \gamma_5 q_\mu
    \right.\nonumber\\
    && -g_1\left[M_{B_Q}+M_{B_Q^*}
\right]\!\not\!{q}\!\not\!{p} \gamma_5\epsilon_\mu
+g_2\!\not\!{q}\!\not\!{p} \gamma_5 p\cdot\epsilon
q_\mu-g_2\!\not\!{q}\!\not\!{p} \gamma_5 q \cdot p \epsilon_\mu
\nonumber\\
&&\left.-\frac{g_2}{2}\!\not\!{q}\!\not\!{p} \gamma_5 q^2
\epsilon_\mu-g_3 \!\not\!{q}\!\not\!{p} \gamma_5 q^2
\epsilon_\mu+\cdots\right\}\left[1/\frac{1}{\sqrt{2}}\right]+\cdots
\, ,
\end{eqnarray}
where the following definitions have been used,
\begin{eqnarray}
\langle 0| J (0)|B_Q(p)\rangle &=&\lambda_{B_Q} U(p,s) \, , \nonumber \\
\langle 0| J_\mu (0)|B_Q^*(p)\rangle &=&\lambda_{B_Q^*} U_\mu(p,s) \, , \nonumber \\
\sum_sU(p,s)\overline {U}(p,s)&=&\!\not\!{p}+M_{B_Q} \, , \nonumber \\
\sum_s U_\mu(p,s) \overline{U}_\nu(p,s)
&=&-(\!\not\!{p}+M_{B_Q^*})\left( g_{\mu\nu}-\frac{\gamma_\mu
\gamma_\nu}{3}-\frac{2p_\mu p_\nu}{3M_{B_Q^*}^2}+\frac{p_\mu
\gamma_\nu-p_\nu \gamma_\mu}{3M_{B_Q^*}} \right) \,  ,
\end{eqnarray}
the factors $1$ and $\frac{1}{\sqrt{2}}$ correspond to the
correlation functions $\Pi_{\mu}^{\phi}(p,q)$ and
$\Pi_{\mu}^{\rho_0}(p,q)$ respectively.
 The current $J_\mu(x)$ couples
not only to the  spin-parity $J^P=\frac{3}{2}^+$ states, but also to
the   spin-parity $J^P=\frac{1}{2}^-$ states.
 For a generic $\frac{1}{2}^-$ resonance   $\widetilde{B}_Q^*$,
$ \langle0|J_{\mu}(0)|\widetilde{B}_Q^*(p)\rangle=\lambda_{*}
 (\gamma_{\mu}-4\frac{p_{\mu}}{M_{*}})U^{*}(p,s)$, where $\lambda^{*}$ is  the  pole residue, $M_{*}$ is the
mass, and the spinor $U^*(p,s)$  satisfies the usual Dirac equation
$(\not\!\!p-M_{*})U^{*}(p)=0$.
 In this article, we choose the tensor structures $\!\not\!{\epsilon}\!\not\!{p} \gamma_5 q_\mu$,
$\!\not\!{q}\!\not\!{p} \gamma_5 p\cdot\epsilon q_\mu$ and
$\!\not\!{q}\!\not\!{p} \gamma_5  \epsilon_\mu$, the baryon state
$\widetilde{B}_Q^*$  has no contamination, for example, we can study
the contribution of the $\frac{1}{2}^-$ baryon state
$\widetilde{B}_Q^*$ to the correlation functions
$\Pi_{\mu}^{\phi}(p,q)$,
\begin{eqnarray}
\Pi_{\mu}^{\phi}(p,q)&=&\frac{\langle0| J(0)| B_Q(q+p)\rangle\langle
B_Q(q+p)| \widetilde{B}_Q^*(p) \phi(q) \rangle \langle
\widetilde{B}_Q^*(p)|\bar{J}_\mu(0)| 0\rangle}
  {\left[M_{B_Q}^2-(q+p)^2\right]\left[M_{*}^2-p^2\right]}  + \cdots \nonumber \\
&=&\lambda_{B_Q}\lambda_{*} \frac{\!\not\!{q}+\!\not\!{p}+M_{B_Q}}
{M_{B_Q}^2-(q+p)^2} \left[g_V \!\not\!{\epsilon}
    +ig_T \frac{ \epsilon^\alpha \sigma_{\alpha\beta}
    q^\beta}{M_{B_Q}+M_*}\right]\gamma_5
    \frac{\!\not\!{p}+M_{*}}{M_{*}^2-p^2}\left[\gamma_{\mu}-4\frac{p_{\mu}}{M_{*}}\right]
\nonumber\\
&&+\cdots\nonumber\\
&=&f_1(\gamma,p,q,\epsilon)\gamma_\mu+f_2(\gamma,p,q,\epsilon)p_\mu
+\cdots \, ,
\end{eqnarray}
where we introduce the strong coupling constants $g_V$ and $g_T$ to
parameterize the vertexes  $\langle B_Q(q+p)| \widetilde{B}_Q^*(p)
\phi(q) \rangle$, the notations $f_1$ and $f_2$ are functions of
$\gamma_\alpha$, $\gamma_5$, $\epsilon_\alpha$, $p_\alpha$ and
$q_\alpha$, here we order the Dirac matrixes  as
$\!\not\!{\epsilon}, \, \!\not\!{q},\, \!\not\!{p}, \,\gamma_5$.

In the following, we briefly outline the  operator product expansion
for the correlation functions  $\Pi_{\mu }(p,q)$  in perturbative
QCD. The calculations are performed at the large space-like momentum
regions $(q+p)^2\ll 0$  and $p^2\ll 0$, which correspond to the
small light-cone distance $x^2\approx 0$ required by the validity of
the operator product expansion approach. We write down the "full"
propagator of a massive    quark in the presence of the quark and
gluon condensates firstly \cite{LCSR89,PRT85},
\begin{eqnarray}
S_{ij}(x)&=& \frac{i\delta_{ij}\!\not\!{x}}{ 2\pi^2x^4}
-\frac{\delta_{ij}m_s}{4\pi^2x^2}-\frac{\delta_{ij}}{12}\langle
\bar{s}s\rangle +\frac{i\delta_{ij}}{48}m_s
\langle\bar{s}s\rangle\!\not\!{x}-\frac{\delta_{ij}x^2}{192}\langle \bar{s}g_s\sigma Gs\rangle\nonumber\\
&& +\frac{i\delta_{ij}x^2}{1152}m_s\langle \bar{s}g_s\sigma
 Gs\rangle \!\not\!{x}-\frac{i}{16\pi^2x^2} \int_0^1 dv G^{ij}_{\mu\nu}(vx) \left[(1-v)\!\not\!{x}
\sigma^{\mu\nu}+v\sigma^{\mu\nu} \!\not\!{x}\right]  +\cdots \, ,\nonumber\\
S_Q^{ij}(x)&=&\frac{i}{(2\pi)^4}\int d^4k e^{-ik \cdot x} \left\{
\frac{\delta_{ij}}{\!\not\!{k}-m_Q}
-\frac{g_sG^{\alpha\beta}_{ij}}{4}\frac{\sigma_{\alpha\beta}(\!\not\!{k}+m_Q)+(\!\not\!{k}+m_Q)
\sigma_{\alpha\beta}}{(k^2-m_Q^2)^2}\right.\nonumber\\
&&\left.+\frac{\pi^2}{3} \langle \frac{\alpha_sGG}{\pi}\rangle
\delta_{ij}m_Q \frac{k^2+m_Q\!\not\!{k}}{(k^2-m_Q^2)^4}
+\cdots\right\} \, ,
\end{eqnarray}
where $\langle \bar{s}g_s\sigma Gs\rangle=\langle
\bar{s}g_s\sigma_{\alpha\beta} G^{\alpha\beta}s\rangle$  and
$\langle \frac{\alpha_sGG}{\pi}\rangle=\langle
\frac{\alpha_sG_{\alpha\beta}G^{\alpha\beta}}{\pi}\rangle$ (the
corresponding full propagators $U_{ij}(x)$ and $D_{ij}(x)$ of the
quarks $u$ and $d$  respectively can be obtained with a simple
replacement), then contract the quark fields in the correlation
functions $\Pi_\mu(p,q)$ with Wick theorem, and obtain the following
results:
\begin{eqnarray}
\Pi_\mu^{\Xi_Q^*\Xi_Q\phi}(p,q)&=&i\epsilon^{ijk}\epsilon^{i'j'k'}
\int d^4x e^{-i
p\cdot x} \nonumber \\
&&\gamma_5\gamma^\alpha S_Q^{kk'}(-x) Tr\left[\gamma_\alpha \langle
0|s_j(0)\bar{s}_{j'}(x) |\phi(q)\rangle \gamma_\mu
CU/D_{ii'}^T(-x)C\right]\, ,
\end{eqnarray}

\begin{eqnarray}
\Pi_\mu^{\Xi_Q^*\Xi_Q\rho}(p,q)&=&i\epsilon^{ijk}\epsilon^{i'j'k'}
\int d^4x e^{-i p
\cdot x} \nonumber \\
 &&\gamma_5\gamma^\alpha S_Q^{kk'}(-x)Tr\left[ \gamma_\alpha S_{jj'}(-x)\gamma_\mu C\langle 0|q_i(0)\bar{q}_{i'}(x) |\rho(q)\rangle^T C\right]\,
 ,
\end{eqnarray}

\begin{eqnarray}
\Pi_\mu^{\Sigma_Q^*\Sigma_Q\rho}(p,q)&=&Ai\epsilon^{ijk}\epsilon^{i'j'k'}
\int d^4x e^{-i p
\cdot x} \nonumber \\
&&\left\{\varpi\gamma_5\gamma^\alpha S_Q^{kk'}(-x)
Tr\left[\gamma_\alpha \langle 0|u_j(0)\bar{u}_{j'}(x)
|\rho(q)\rangle \gamma_\mu CU_{ii'}^T(-x)C\right]
 \right.\nonumber \\
 &&\left. +\gamma_5\gamma^\alpha S_Q^{kk'}(-x)Tr\left[ \gamma_\alpha U_{jj'}(-x)\gamma_\mu C\langle 0|u_i(0)\bar{u}_{i'}(x) |\rho(q)\rangle^T C\right]\right\}\,
 ,
\end{eqnarray}
here we take isospin limit for the  quarks $u$ and $d$, the symmetry
factor $A=1$, $\varpi=-1$ for the channels
$\Sigma^{*}(Qud)\Sigma(Qud)\rho$, and $A=\pm2$ and $\varpi=1$ for
the channels $\Sigma^{*}(Quu)\Sigma(Quu)\rho$ and
$\Sigma^{*}(Qdd)\Sigma(Qdd)\rho$ respectively.

Performing the  Fierz re-ordering to extract the contributions from
the two-particle and three-particle vector meson light-cone
distribution amplitudes respectively,
 then   substituting  the full $q$  and $Q$ quark propagators into the
correlation functions in Eqs.(9-11) and completing   the integral in
the coordinate space, finally  integrating  over the variable $k$,
we can obtain the correlation functions $\Pi_\mu(p,q)$ at the level
of quark-gluon degree of freedom.
 In calculation, the two-particle and three-particle vector meson
light-cone distribution amplitudes have been used
\cite{VMLC981,VMLC982,VMLC2003,VMLC2007}. The parameters in the
light-cone distribution amplitudes are scale dependent and are
estimated with the QCD sum rules \cite{VMLC2003,VMLC2007}. In this
article, the energy scale $\mu$ is chosen to be $\mu=1\,\rm{GeV}$.

Taking double Borel transform  with respect to the variables
$Q_1^2=-p^2$ and $Q_2^2=-(p+q)^2$ respectively,  then subtracting
the contributions from the high resonances and continuum states by
introducing  the threshold parameter $s_0$ (i.e. $ M^{2n}\rightarrow
\frac{1}{\Gamma[n]}\int_0^{s_0} ds s^{n-1}e^{-\frac{s}{M^2}}$),
finally we can obtain 30  sum rules  for the strong coupling
constants $g_{1}$, $g_{2}$ and
$\rm{G}3=-\left(M_{B_Q}+M_{B_Q}^*\right)g_1-M_{\phi/\rho_0}^2
\left(\frac{g_2}{2}+g_3\right)$ respectively, the explicit
expressions are presented in the appendix A\footnote{Here we present
some technical details in calculating the correlation functions
$\Pi_\mu^{\Xi_Q^*\Xi_Q\phi}(p,q)$ to illustrate the procedure,
\begin{eqnarray}
\Pi_\mu(p,q)&=&\frac{i\epsilon^{ijk}\epsilon^{ij'k'}}{12} \int d^4x
e^{-i p\cdot x} \langle 0|\bar{s}(x)\gamma^\lambda \gamma_5s(0)
|\phi(q)\rangle \gamma_5\gamma^\alpha S_Q^{kk'}(-x)
Tr\left[\gamma_\alpha
\gamma_\lambda\gamma_5 \gamma_\mu CU_{jj'}^T(-x)C\right]+\cdots\nonumber \\
 &=&\frac{3\widetilde{f}_\phi
M_\phi}{32\pi^6}\gamma_5\gamma^\alpha\left[\epsilon_\mu
q_\alpha-\epsilon_\alpha q_\mu \right] \int_0^1du
g_{\perp}^{(a)}(1-u) \int d^Dxd^Dke^{i
(k-p-uq)\cdot x}\frac{\!\not\!{k}}{k^2-m_Q^2}\frac{1}{x^2}+\cdots\nonumber \\
&=&\left[\!\not\!{q}\!\not\!{p}\gamma_5\epsilon_\mu-\!\not\!{\epsilon}\!\not\!{p}\gamma_5
q_\mu\right] \frac{\widetilde{f}_\phi M_\phi}{16\pi^2} \int_0^1du
g_{\perp}^{(a)}(1-u)\int_0^1 dt t \frac{\Gamma(\varepsilon)}{\left[(p+uq)^2-\widetilde{m}_Q^2\right]^\varepsilon}\mid_{\varepsilon\rightarrow0}+\cdots\nonumber \\
&\longrightarrow&\left[\!\not\!{q}\!\not\!{p}\gamma_5\epsilon_\mu-\!\not\!{\epsilon}\!\not\!{p}\gamma_5
q_\mu\right] \frac{\widetilde{f}_\phi M_\phi}{16\pi^2} \int_0^1du
g_{\perp}^{(a)}(1-u)\int_0^1 dt t \nonumber \\
&&\frac{M^4}{M_1^2M_2^2}\exp\left[-\frac{\widetilde{m}_Q^2+u(1-u)M_\phi^2}{M^2}
\right]\delta(u-u_0)+\cdots \,\, ({\bf \rm{take \,\, double\,\, Borel\,\, transform}})\nonumber \\
&\longrightarrow&\left[\!\not\!{q}\!\not\!{p}\gamma_5\epsilon_\mu-\!\not\!{\epsilon}\!\not\!{p}\gamma_5
q_\mu\right] \frac{M^4E_1(x)}{M_1^2M_2^2}\frac{\widetilde{f}_\phi
M_\phi g_{\perp}^{(a)}(1-u_0)}{16\pi^2}
\int_0^1 dt t \nonumber \\
&&\exp\left[-\frac{\widetilde{m}_Q^2+u_0(1-u_0)M_\phi^2}{M^2}
\right]+\cdots \,\, ({\bf \rm{subtract  \,\, continuum\,\, contributions}})\nonumber \\
&=&\frac{\lambda_{\Xi'_Q}\lambda_{\Xi_Q^*}}{M_1^2M_2^2}\exp\left[-\frac{M_{\Xi_Q^*}^2}{M_1^2}-\frac{M_{\Xi'_Q}^2}{M_2^2}\right]
\left\{g_1 \left[M_{\Xi'_Q}+M_{\Xi_Q^*}
\right]\!\not\!{\epsilon}\!\not\!{p} \gamma_5 q_\mu
    \right.\nonumber\\
    && \left.-g_1\left[M_{\Xi'_Q}+M_{\Xi_Q^*}
\right]\!\not\!{q}\!\not\!{p} \gamma_5\epsilon_\mu
+g_2\!\not\!{q}\!\not\!{p} \gamma_5 p\cdot\epsilon
q_\mu-g_2\!\not\!{q}\!\not\!{p} \gamma_5 q \cdot p \epsilon_\mu
+\cdots\right\}+\cdots \, .
\end{eqnarray}
For technical details about the Borel transform, one can consult the
excellent  review \cite{LCSRreview}. }.

\section{Numerical result and discussion}
The masses of the established   hadrons are taken from the Particle
Data Group $M_\phi=1.019455\,\rm{GeV}$, $M_\rho=0.77549\,\rm{GeV}$,
$M_\omega=0.78265\,\rm{GeV}$, $M_{\Xi_c^*}=2.6459\,\rm{GeV}$,
$M_{\Sigma_c^{*++}}=2.5184\,\rm{GeV}$,
 $M_{\Sigma_c^{*+}}=2.5175\,\rm{GeV}$,
 $M_{\Sigma_c^{*0}}=2.5180\,\rm{GeV}$, $M_{\Sigma_b^{*+}}=5.8290\,\rm{GeV}$,
 $M_{\Sigma_b^{*-}}=5.8364\,\rm{GeV}$,
$M_{\Xi'^+_c}=2.5756\,\rm{GeV}$, $M_{\Xi'^0_c}=2.5779\,\rm{GeV}$,
  $M_{\Sigma_c^{++}}=2.45402\,\rm{GeV}$,
 $M_{\Sigma_c^{+}}=2.4529\,\rm{GeV}$,
 $M_{\Sigma_c^{0}}=2.45376\,\rm{GeV}$,
  $M_{\Sigma_b^{+}}=5.8078\,\rm{GeV}$,
 and $M_{\Sigma_b^{-}}=5.8152\,\rm{GeV}$
\cite{PDG}.  In calculation, we take the average values of the
masses in each isospin  multiplet and neglect the small isospin
splitting in the heavy baryon  multiplet.

The parameters which determine the vector meson light-cone
distribution amplitudes are $f_\phi=(0.215\pm0.005)\,\rm{GeV}$,
$f_\phi^{\perp}=(0.186\pm0.009)\,\rm{GeV}$, $a_1^{\parallel}=0.0$,
$a_1^{\perp}=0.0$, $a_2^{\parallel}=0.18\pm0.08$,
$a_2^{\perp}=0.14\pm0.07$, $\zeta^{\parallel}_3=0.024\pm 0.008$,
$\widetilde{\lambda}_3^{\parallel}=0.0$,
$\widetilde{\omega}_3^{\parallel}=-0.045\pm 0.015$,
$\kappa_3^{\parallel}=0.0$, $\omega_3^\parallel=0.09\pm0.03$,
$\lambda_3^\parallel=0.0$, $\kappa_3^\perp=0.0$,
$\omega_3^\perp=0.20\pm0.08$, $\lambda_3^\perp=0.0$,
$\varsigma_4^\parallel=0.00\pm 0.02$,
$\widetilde{\omega}_4^\parallel=-0.02\pm0.01$,
$\varsigma_4^\perp=-0.01\pm 0.03$,
$\widetilde{\varsigma}_4^\perp=-0.03\pm 0.04$,
$\kappa_4^\parallel=0.0$, $\kappa_4^\perp=0.0$ for the $\phi$-meson;
and $f_\rho=(0.216\pm0.003)\,\rm{GeV}$,
$f_\rho^{\perp}=(0.165\pm0.009)\,\rm{GeV}$, $a_1^{\parallel}=0.0$,
$a_1^{\perp}=0.0$, $a_2^{\parallel}=0.15\pm0.07$,
$a_2^{\perp}=0.14\pm0.06$, $\zeta^{\parallel}_3=0.030\pm 0.010$,
$\widetilde{\lambda}_3^{\parallel}=0.0$,
$\widetilde{\omega}_3^{\parallel}=-0.09\pm 0.03$,
$\kappa_3^{\parallel}=0.0$, $\omega_3^\parallel=0.15\pm0.05$,
$\lambda_3^\parallel=0.0$, $\kappa_3^\perp=0.0$,
$\omega_3^\perp=0.55\pm0.25$, $\lambda_3^\perp=0.0$,
$\varsigma_4^\parallel=0.07\pm 0.03$,
$\widetilde{\omega}_4^\parallel=-0.03\pm0.01$,
$\varsigma_4^\perp=-0.03\pm 0.05$,
$\widetilde{\varsigma}_4^\perp=-0.08\pm 0.05$,
$\kappa_4^\parallel=0.0$, and $\kappa_4^\perp=0.0$ for the
$\rho$-meson  at the energy scale $\mu=1\, \rm{GeV}$
\cite{VMLC2003,VMLC2007}.

The QCD input parameters are taken to be the standard values
$m_s=(140\pm 10 )\,\rm{MeV}$, $m_c=(1.35\pm 0.10)\,\rm{GeV}$,
$m_b=(4.7\pm 0.1)\,\rm{GeV}$,
 $\langle \bar{q}q \rangle=-(0.24\pm
0.01 \,\rm{GeV})^3$, $\langle \bar{s}s \rangle=(0.8\pm 0.2 )\langle
\bar{q}q \rangle$, $\langle \bar{s}g_s\sigma Gs \rangle=m_0^2\langle
\bar{s}s \rangle$, $\langle \bar{q}g_s\sigma Gq \rangle=m_0^2\langle
\bar{q}q \rangle$, $m_0^2=(0.8 \pm 0.2)\,\rm{GeV}^2$, and $\langle
\frac{\alpha_s GG}{\pi}\rangle=(0.33\,\rm{GeV})^4 $  at the energy
scale $\mu=1\, \rm{GeV}$ \cite{SVZ79,PRT85,Ioffe2005}.

The bottom baryon states $\Xi^*_b$ and $\Xi'_b$ have not been
observed yet, we study their masses with the conventional QCD sum
rules. The masses $M_{B_Q}$ and $M_{B^*_Q}$ and pole residues
$\lambda_{B_Q}$ and $\lambda_{B^*_Q}$ are determined by the
following correlation functions,
\begin{eqnarray}
\Pi_{\mu \nu}(p)&=&i \int d^4x \, e^{i p \cdot x} \,
\langle 0 |T\left\{ J_\mu(x)\bar{J}_\nu(0)\right\}|0\rangle \, , \nonumber\\
\Pi(p)&=&i \int d^4x \, e^{i p \cdot x} \,
\langle 0 |T\left\{ J(x)\bar{J}(0)\right\}|0\rangle \, , \\
J^\Xi(x)&=& \epsilon^{ijk}  u^T_i(x)C\gamma_\mu s_j(x) \gamma_5 \gamma^\mu Q_k(x)  \, ,  \nonumber \\
J^\Sigma(x)&=& \epsilon^{ijk}  u^T_i(x)C\gamma_\mu d_j(x) \gamma_5 \gamma^\mu Q_k(x)  \, ,  \nonumber \\
J^\Xi_\mu(x)&=& \epsilon^{ijk} u^T_i(x)C\gamma_\mu s_j(x)  Q_k(x)\, ,  \nonumber \\
J^\Sigma_\mu(x)&=& \epsilon^{ijk}  u^T_i(x)C\gamma_\mu d_j(x)
Q_k(x)\, .
\end{eqnarray}

In Refs.\cite{Huang08,Narison09}, the masses of the heavy baryon
states containing  one heavy quark are studied using the QCD sum
rules, the pole residues are not calculated. In this article, we
take the simple Ioffe type interpolating currents,  which are
constructed by considering  the  diquark  theory  and the heavy
quark symmetry \cite{Jaffe2003,Jaffe2004}.
 We  insert  a complete set of
intermediate baryon states with the same quantum numbers as the
current operators $J(x)$ and $J_\mu$ into the correlation functions
$\Pi(p)$ and $\Pi_{\mu\nu}(p)$ to obtain the hadronic representation
\cite{SVZ79,PRT85}. After isolating the pole terms of the lowest
states $\Xi_Q^*$, $\Xi'_Q$, $\Sigma^*_Q$ and $\Sigma_Q$, we obtain
the following results:
\begin{eqnarray}
\Pi_{\mu\nu}(p)&=&\lambda_{B_Q^*}^2\frac{M_{B_Q^*}+\!\not\!{p}}{M_{B_Q^*}^2-p^2}\left[-g_{\mu\nu}+\cdots\right]
+\cdots \, \, ,\nonumber \\
\Pi(p)&=&\lambda_{B_Q}^2\frac{M_{B_Q}+\!\not\!{p}}{M_{B_Q}^2-p^2}
+\cdots \, \, ,
\end{eqnarray}
we choose the tensor structures $g_{\mu\nu}$,
$\!\not\!{p}g_{\mu\nu}$, $1$ and $\!\not\!{p}$ for analysis. After
performing the standard procedure  of the QCD sum rules, we obtain
sixteen  sum rules for the heavy baryons states $B_Q^*$ and $B_Q$,
\begin{eqnarray}
\lambda_{i}^2 e^{-\frac{M_i^2}{M^2}}= \int_{\Delta_i}^{s^0_i} ds
\rho_i^{A}(s)e^{-\frac{s}{M^2}} \, , \nonumber \\
\lambda_{i}^2 M_ie^{-\frac{M_i^2}{M^2}}= \int_{\Delta_i}^{s^0_i} ds
\rho_i^{B}(s)e^{-\frac{s}{M^2}} \, ,
\end{eqnarray}
where the $i$ denote the channels $\Xi_Q^*$,
   $\Xi'_Q$, $\Sigma_Q^*$ and  $\Sigma_Q$
     respectively;
      the $s_i^0$ are the corresponding continuum threshold parameters and the $M^2$ is the Borel
      parameter.
 The thresholds $\Delta_i$ can be sorted into two sets,  we introduce the $q\bar{q}$,
$q\bar{s}$ to denote the light quark constituents in the heavy
baryon states to simplify the notations, $\Delta_{q\bar{q}}=m_Q^2$,
$\Delta_{q\bar{s}}=(m_Q+m_s)^2$. The explicit expressions of the
spectral densities $\rho_i^A(s)$ and $\rho^B_i(s)$ are given in the
appendix B.

Differentiate  the Eq.(16) with respect to  $\frac{1}{M^2}$, then
eliminate the
 pole residues $\lambda_{i}$, we can obtain the sixteen  sum rules for
 the masses  of the heavy baryon states $B_Q^*$ and $B_Q$,
 \begin{eqnarray}
 M_i^2= \frac{\int_{\Delta_i}^{s^0_i} ds
s\rho_i^{A/B}(s)e^{-\frac{s}{M^2}} }{\int_{\Delta_i}^{s^0_i} ds
\rho_i^{A/B}(s)e^{-\frac{s}{M^2}}}\, .
\end{eqnarray}

In the conventional QCD sum rules \cite{SVZ79,PRT85}, there are two
criteria (pole dominance and convergence of the operator product
expansion) for choosing  the Borel parameter $M^2$ and threshold
parameter $s_0$.  We impose the two criteria on the heavy baryon
states to choose the Borel parameter $M^2$ and threshold parameter
$s_0$, the values are shown in Table 1. Finally we obtain the values
of the masses and pole resides of
 the  heavy baryon states $B_Q^*$ and $B_Q$, which are  shown in Table 2.

 From Table 2, we can see that the average values of the masses with
 the tensor structures $\!\not\!{p}g_{\mu\nu}$ ($\!\not\!{p}$) and $g_{\mu\nu}$ ($1$)
 can reproduce the experimental data
approximately for the established heavy baryon states.  So it is
reasonable to take the average values $M_{\Xi_b^*}=5.98\,\rm{GeV}$
and $M_{\Xi'_b}=5.95\,\rm{GeV}$ for the un-established bottom baryon
states $\Xi_b^*$ and $\Xi'_b$ in numerical analysis. The values of
the pole residues from different tensor structures differ greatly
from each other in some channels, for example, $\Xi_b'$, $\Sigma_b$,
$\Xi_c'$, $\Sigma_c$. In this article, we take the average values
and assume uniform uncertainties (about $20\%$) for the pole
residues in all channels, the   uncertainties originate from the
parameters other than the Borel parameter $M^2$ are about $20\%$, we
subtract the uncertainties originate from the Borel parameter from
the total uncertainties to avoid double counting. The values of the
pole residues are
$\lambda_{\Xi^*_b}=(4.5\pm0.8)\times10^{-2}\,\rm{GeV}^3$,
$\lambda_{\Xi^*_c}=(3.1\pm0.5)\times10^{-2}\,\rm{GeV}^3$,
$\lambda_{\Xi'_b}=(7.5\pm1.5)\times10^{-2}\,\rm{GeV}^3$,
$\lambda_{\Xi'_c}=(5.3\pm1.0)\times10^{-2}\,\rm{GeV}^3$,
$\lambda_{\Sigma^*_b}=(3.8\pm0.7)\times10^{-2}\,\rm{GeV}^3$,
$\lambda_{\Sigma^*_c}=(2.7\pm0.5)\times10^{-2}\,\rm{GeV}^3$,
$\lambda_{\Sigma_b}=(6.5\pm1.2)\times10^{-2}\,\rm{GeV}^3$, and
$\lambda_{\Sigma_c}=(4.5\pm0.9)\times10^{-2}\,\rm{GeV}^3$. The
threshold parameters  are taken as $s_0=(11.0\pm1.0)\,\rm{GeV}^2$,
$(10.5\pm1.0)\,\rm{GeV}^2$, $(45.0\pm1.0)\,\rm{GeV}^2$ and
$(46.0\pm1.0)\,\rm{GeV}^2$ in the channels $\Xi_c^*$($\Xi'_c$),
$\Sigma_c^*$($\Sigma_c$), $\Xi_b^*$($\Xi'_b$) and
$\Sigma_b^*$($\Sigma_b$)  respectively;  the Borel parameters are
taken as $M^2=(2.0-3.0)\,\rm{GeV^2}$ and $(5.0-6.0)\,\rm{GeV^2}$ in
the charm  and bottom channels respectively. Those parameters  are
determined by the two-point QCD sum rules  to avoid possible
contaminations from the high resonances and continuum states. In
calculation, we observe that the values of the strong coupling
constants $g_1$, $g_2$ and G3 are insensitive to threshold
parameters $s_0$.

\begin{table}
\begin{center}
\begin{tabular}{|c|c|c|c|}
\hline\hline & $M^2 (\rm{GeV}^2)$& $s_0 (\rm{GeV}^2)$\\ \hline
      $\Xi^*_b$  &$5.0-6.0$ &$46.0\pm0.5$\\ \hline
      $\Xi'_b$  &$4.8-5.6$ &$44.5\pm0.5$\\ \hline
       $\Sigma^*_b$  &$5.0-6.0$ &$45.0\pm0.5$\\ \hline
        $\Sigma_b$  &$4.8-5.6$ &$43.5\pm0.5$\\ \hline
         $\Xi^*_c$  &$2.0-3.0$ &$11.0\pm0.5$\\ \hline
      $\Xi'_c$  &$2.0-2.8$ &$10.5\pm0.5$\\ \hline
       $\Sigma^*_c$  &$2.0-3.0$ &$10.5\pm0.5$\\ \hline
        $\Sigma_c$  &$1.9-2.7$ &$10.0\pm0.5$\\ \hline
    \hline
\end{tabular}
\end{center}
\caption{ The Borel parameters $M^2$ and threshold parameters $s_0$
for the heavy baryon states.}
\end{table}

\begin{table}
\begin{center}
\begin{tabular}{|c|c|c|c|c|}
\hline\hline &$\!\not\!{p}g_{\mu\nu}/\!\not\!{p}\,\,(M)
$&$g_{\mu\nu}/1\,\,(M) $&
$\!\not\!{p}g_{\mu\nu}/\!\not\!{p}\,\,(\lambda)$&$g_{\mu\nu}/1\, \,(\lambda) $\\
\hline $\Xi_b^*  $&  $6.04\pm0.14$ & $5.92\pm0.20$ & $4.8\pm 1.0$ &$4.2\pm 1.3$ \\
\hline $\Xi'_b $&  $6.04\pm0.10$ & $5.85\pm0.15$ & $9.0\pm 1.8$ &$6.0\pm 1.3$ \\
 \hline $\Sigma_b^*  $& $5.95\pm0.14$  & $5.72\pm0.25$ & $4.2\pm 1.0$ &$3.3\pm 1.3$ \\
\hline $\Sigma_b $& $5.96\pm0.10$  & $5.73\pm0.16$ & $8.0\pm 1.7$ &$5.0\pm 1.2$ \\
\hline $\Xi_c^*  $&  $2.60\pm0.20$ & $2.68\pm0.18$ & $3.1\pm 0.8$ &$3.1\pm 0.7$ \\
\hline $\Xi'_c $&  $2.65\pm0.14$ & $2.54\pm 0.17$ & $6.2\pm 1.5$ &$4.4\pm 0.9$ \\
\hline $\Sigma_c^*  $& $2.50\pm0.20$  & $2.59\pm0.19$ & $2.7\pm 0.7$ &$2.7\pm 0.7$ \\
\hline $\Sigma_c $&  $2.54\pm0.15$ & $2.42\pm0.20$  & $5.4\pm 1.4$ &$3.6\pm 1.0$ \\
         \hline \hline
\end{tabular}
\end{center}
\caption{ The masses and pole residues of the heavy baryon states
from the sum rules with different tensor structures. The masses $M$
are in unit of $\rm{GeV}$ and the pole residues $\lambda$ are in
unit of $10^{-2}\rm{GeV}^3$.}
\end{table}

 The main uncertainties originate
from the parameters $\lambda_{B_Q}$, $\lambda_{B_Q^*}$ (as the
strong coupling constants $g_1$, $g_2$ and G3 $\propto
{1\over\lambda_{B_Q} \lambda_{B_Q^*}}$) and $m_{Q}$,
  the variations of those parameters can lead to relatively large
changes for the numerical values, and almost saturate the total
uncertainties, i.e. the variations of the two hadronic parameters
$\lambda_{B_Q}$ and $\lambda_{B_Q^*}$ lead to an uncertainty about
$20\%\times \sqrt{2}=28\%$, and the variations of the   $m_Q$ lead
to an uncertainty about $(10-20)\%$, refining those parameters is of
great importance. In the case of the sum rules for the strong
coupling constants $g_2$, the values are not stable enough with
variations of the Borel parameter, additional uncertainties  are
introduced, the total uncertainties  are very large, see Table 3.
The contributions from the strong coupling constants $g_2$ to the
radiative decay widths are very small comparing with the
corresponding ones from  the $g_1$, the predictions are insensitive
to the Borel parameter. Although there are many parameters in the
light-cone distributions amplitudes \cite{VMLC2003,VMLC2007}, the
uncertainties originate from those parameters are rather small. In
calculation, we neglect the contributions from the high dimension
vacuum condensates, such as $\langle f_{abc}G^aG^bG^c\rangle$,
$\langle \bar{q}q\rangle\langle\frac{\alpha_sGG}{\pi}\rangle$,
$\langle \bar{s}s\rangle\langle\frac{\alpha_sGG}{\pi}\rangle$, etc.
They are greatly suppressed by the large  numerical denominators and
additional inverse powers of the  Borel parameter $\frac{1}{M^{2}}$,
and would not play any significant roles. Furthermore, we neglect
some terms involving  the light-cone distributions amplitudes
$\widetilde{f}(\bar{u}_0)$ and
$\widetilde{\widetilde{f}}(\bar{u}_0)$ in case of the contributions
from the   terms  $f(\bar{u}_0)$ are small, as
\begin{eqnarray}
\frac{\widetilde{f}(\bar{u}_0)}{f(\bar{u}_0)}\approx 40\%,
 \, \, \, \, \, \, \frac{\widetilde{\widetilde{f}}(\bar{u}_0)}{f(\bar{u}_0)}\approx
10\% \, .
\end{eqnarray}

Taking into account all the uncertainties of the revelent
parameters, finally we obtain the numerical results of the strong
coupling constants $g_1$, $g_2$ and $\rm{G}3$, which are shown in
the   Table 3. We estimate  the uncertainties $\delta$  with the
formula $\delta=\sqrt{\sum_i\left(\frac{\partial f}{\partial
x_i}\right)^2\mid_{x_i=\bar{x}_i} (x_i-\bar{x}_i)^2}$,
 where the $f$ denote  strong coupling constants $g_1$, $g_2$ and
$\rm{G}3$,  the $x_i$ denote the revelent     parameters $m_Q$,
$\langle \bar{q}q \rangle$, $\langle \bar{s}s \rangle$, $\cdots$. In
the numerical calculations, we take the  approximation
$\left(\frac{\partial f}{\partial x_i}\right)^2
(x_i-\bar{x}_i)^2\approx \left[f(\bar{x}_i\pm \Delta
x_i)-f(\bar{x}_i)\right]^2$ for simplicity.  For the central values
of the strong coupling constants,
$\frac{g_1^{B^*_bB_bV}}{g_1^{B^*_cB_cV}}\approx (70-80)\%$,
$\frac{{\rm{G3}}_{B^*_bB_bV}(M_{B_c^*}+M_{B_c})}{{\rm{G3}}_{B^*_cB_cV}(M_{B_b^*}+M_{B_b})}\approx
(80-90)\%$, the heavy quark symmetry works rather well. Those strong
coupling constants  in the vertexes $B_Q^*V_Q V$ are basic
parameters in describing the interactions among the heavy baryon
states, once reasonable values are obtained, we can use them to
perform phenomenological analysis.

\begin{table}
\begin{center}
\begin{tabular}{|c|c|c|c|c|}
\hline\hline Vertexes  & $-g_1 (\rm{GeV}^{-1})$ & $-g_2 (\rm{GeV}^{-2})$&$\rm{G3}$\\
\hline
      $\Xi_c^{*+}{\Xi'}_c^+\phi$, \,$\Xi_c^{*0}{\Xi'}_c^0\phi$  &$2.98^{+1.18}_{-0.81}$ &$0.62^{+0.63}_{-0.31}$&$9.75^{+3.87}_{-2.68}$\\ \hline
       $\Xi_c^{*+}{\Xi'}_c^+\rho$,\,$\Xi_c^{*+}{\Xi'}_c^+\omega$,\,$\Xi_c^{*0}{\Xi'}_c^0\omega$& $3.54^{+1.39}_{-0.97}$ & $0.47^{+0.46}_{-0.24}$&$13.61^{+5.42}_{-3.76}$\\     \hline
      $\Xi_c^{*0}{\Xi'}_c^0\rho_0$ &$-(3.54^{+1.39}_{-0.97})$ & $-(0.47^{+0.46}_{-0.24})$&$-(13.61^{+5.42}_{-3.76})$\\     \hline
         $\Sigma_c^{*++}\Sigma_c^{++}\rho_0$,\,$\Sigma_c^{*++}\Sigma_c^{++}\omega$,\,$\Sigma_c^{*0}\Sigma_c^{0}\omega$,\,$\Sigma_c^{*+}\Sigma_c^{+}\omega$  &$7.07^{+3.09}_{-2.17}$ &$0.97^{+0.87}_{-0.49}$&$25.00^{+10.54}_{-7.53}$\\ \hline
     $\Sigma_c^{*+}\Sigma_c^{+}\rho_0$  &0 &0&0\\ \hline
      $\Sigma_c^{*0}\Sigma_c^{0}\rho_0$  &$-(7.07^{+3.09}_{-2.17})$ &$-(0.97^{+0.87}_{-0.49})$&$-(25.00^{+10.54}_{-7.53})$\\ \hline
             $\Xi_b^{*0}{\Xi'}_b^0\phi$, \,$\Xi_b^{*-}{\Xi'}_b^-\phi$  &$2.25^{+0.93}_{-0.69}$ &$0.11^{+0.16}_{-0.11}$&$20.14^{+8.54}_{-6.12}$\\ \hline
       $\Xi_b^{*0}{\Xi'}_b^0\rho$,\,$\Xi_b^{*0}{\Xi'}_b^0\omega$,\,$\Xi_b^{*-}{\Xi'}_b^-\omega$& $2.73^{+1.14}_{-0.84}$& $0.07^{+0.14}_{-0.05}$&$26.68^{+11.04}_{-8.25}$\\     \hline
      $\Xi_b^{*-}{\Xi'}_b^-\rho$ & $-(2.73^{+1.14}_{-0.84})$& $-(0.07^{+0.14}_{-0.05})$&$-(26.68^{+11.04}_{-8.25})$\\     \hline
    $\Sigma_b^{*+}\Sigma_b^{+}\rho_0$,\,$\Sigma_b^{*+}\Sigma_b^{+}\omega$, \,$\Sigma_b^{*0}\Sigma_b^{0}\omega$,\, $\Sigma_b^{*-}\Sigma_b^{-}\omega$ &$5.05^{+2.17}_{-1.59}$ &$0.13^{+0.27}_{-0.10}$&$47.05^{+19.72}_{-14.71}$\\ \hline
     $\Sigma_b^{*0}\Sigma_b^{0}\rho_0$  &0 &0&0\\ \hline
      $\Sigma_b^{*-}\Sigma_b^{-}\rho_0$ &$-(5.05^{+2.17}_{-1.59})$ &$-(0.13^{+0.27}_{-0.10})$&$-(47.05^{+19.72}_{-14.71})$\\ \hline
           \hline
\end{tabular}
\end{center}
\caption{ The values of the strong coupling constants $g_1$, $g_2$
and G3. }
\end{table}

\begin{table}
\begin{center}
\begin{tabular}{|c|c|c|}
\hline\hline Channels  & $\Gamma$ ($ \rm{KeV}$)\\
\hline
      $\Xi_c^{*+}\to{\Xi'}_c^{+}\gamma $  &$0.96^{+1.47}_{-0.67}$\\ \hline
       $ \Xi_c^{*0}\to{\Xi'}_c^{0}\gamma$ &$1.26^{+0.80}_{-0.46}$\\     \hline
      $\Sigma_c^{*++}\to\Sigma_c^{++}\gamma $ & $6.36^{+6.79}_{-3.31}$\\      \hline
    $\Sigma_c^{*+}\to\Sigma_c^{+}\gamma $ & $0.40^{+0.43}_{-0.21}$\\      \hline
    $\Sigma_c^{*0}\to\Sigma_c^{0}\gamma $ & $1.58^{+1.68}_{-0.82}$\\      \hline
    $\Xi_b^{*0}\to{\Xi'}_b^{0}\gamma $  &$0.047^{+0.077}_{-0.036}$\\ \hline
       $ \Xi_b^{*-}\to{\Xi'}_b^{-}\gamma$ &$0.066^{+0.045}_{-0.027}$\\     \hline
      $\Sigma_b^{*+}\to\Sigma_b^{+}\gamma $ & $0.12^{+0.13}_{-0.06}$\\      \hline
    $\Sigma_b^{*0}\to\Sigma_b^{0}\gamma $ & $0.0076^{+0.0079}_{-0.0040}$\\      \hline
    $\Sigma_b^{*-}\to\Sigma_b^{-}\gamma $ & $0.030^{+0.032}_{-0.016}$\\      \hline
    \hline
\end{tabular}
\end{center}
\caption{ The widths of the  radiative decays $B_Q^* \to B_Q
\gamma$. }
\end{table}

The radiative decays $B_Q^*\to B_Q \gamma$ can be described by the
following electromagnetic lagrangian $\mathcal{L}$,
\begin{eqnarray}
\mathcal{L}&=&-eQ_b \bar{b}\gamma_\mu b A^\mu-eQ_c \bar{c}\gamma_\mu
c A^\mu-eQ_s \bar{s}\gamma_\mu s A^\mu -eQ_u \bar{u}\gamma_\mu u
A^\mu -eQ_d \bar{d}\gamma_\mu d A^\mu\, ,
\end{eqnarray}
where the $A_\mu$ is the electromagnetic field. From the lagrangian
$\mathcal{L}$,  we can obtain the decay amplitudes  with the
assumption of the vector meson dominance, $eT= \langle
B_Q(p)\gamma(q)|\mathcal {L}|B_Q^*(p+q)\rangle $,
\begin{eqnarray}
e T&=&-eQ_s\eta^*_\mu\langle B_Q(p)| \bar{s}\gamma^\mu s
|B_Q^*(p+q)\rangle -eQ_u\eta^*_\mu\langle B_Q(p)|
\bar{u}\gamma^\mu u |B_Q^*(p+q)\rangle\nonumber\\
&&-eQ_d\eta^*_\mu\langle B_Q(p)|
\bar{d}\gamma^\mu d |B_Q^*(p+q)\rangle+\cdots\nonumber \\
&=& -eQ_s\eta^*_\mu f_\phi M_\phi \epsilon_\mu
\frac{i}{q^2-M_\phi^2} \langle\phi(q) B_Q(p)
 |B_Q^*(p+q)\rangle \nonumber \\
&& -eQ_u\eta^*_\mu \frac{1}{\sqrt{2}}f_\rho M_\rho \epsilon_\mu
\frac{i}{q^2-M_\rho^2} \langle\rho(q) B_Q(p)
 |B_Q^*(p+q)\rangle \nonumber \\
 && +eQ_d\eta^*_\mu \frac{1}{\sqrt{2}}f_\rho M_\rho \epsilon_\mu
\frac{i}{q^2-M_\rho^2} \langle\rho(q) B_Q(p)
 |B_Q^*(p+q)\rangle \nonumber \\
 && -eQ_u\eta^*_\mu \frac{1}{\sqrt{2}}f_\omega M_\omega \epsilon_\mu
\frac{i}{q^2-M_\omega^2} \langle\omega(q) B_Q(p)
 |B_Q^*(p+q)\rangle \nonumber \\
 && -eQ_d\eta^*_\mu \frac{1}{\sqrt{2}}f_\omega M_\omega \epsilon_\mu
\frac{i}{q^2-M_\omega^2} \langle\omega(q) B_Q(p)
 |B_Q^*(p+q)\rangle +\cdots \, ,
\end{eqnarray}
 where the
$\eta_\mu$ is the polarization vector of the photon. In the heavy
quark limit, the matrix elements $ \langle B_Q(p)|{\bar Q}\gamma_\mu
Q|B_Q^*(p+q) \rangle\propto M_{J/\psi(\Upsilon)}^{-\frac{3}{2}} $,
and can be neglected, so we consider only the contributions  from
the intermediate vector mesons $\phi(1020)$, $\rho_0(770)$ and
$\omega(782)$. The photon can be viewed as emitted from the light
diquark system while the heavy quark is unaffected by the emission
process.

From the strong coupling constants $g_{1}$ and $g_2$, we can obtain
the decay widths $\Gamma_{B_Q^* \to B_Q \gamma}$,
\begin{eqnarray}
\Gamma_{B_Q^*\to B_Q \gamma}&=&\frac{\alpha
\left(M_{B_Q^*}^2-M_{B_Q}^2\right)}{16 M_{B_Q^*}^3 }
 \sum_{ss'} \mid
T \mid^2 \, ,
\end{eqnarray}
the numerical values are shown in Table 4.

There have been many works  focusing  on the radiative decays of the
$\frac{1}{2}^+$ and $\frac{3}{2}^+$ heavy baryon sextets $B^*_6$ and
$B_6$ to the $\frac{1}{2}^+$ heavy baryon antitriplet $B_{\bar{3}}$,
$B^*_6 \to B_{\bar{3}}\,\gamma$ and $B_6 \to B_{\bar{3}}\,\gamma$,
such as the light-cone QCD sum rules \cite{Aliev0901}, the heavy
hadron chiral perturbation theory \cite{HHCPT-1,HHCPT-2}, the
combination of the heavy quark symmetry and the light diquark
$SU(2N_f)\times O(3)$ symmetry \cite{Heavy-Light}, the relativistic
three-quark model \cite{R3QM-1,R3QM-2}, etc. The works on  the
radiative decays  $B^*_6 \to B_{6}\,\gamma$ are very few, some decay
channels are studied in
 the constituent quark model \cite{CQMwidth} and the non-relativistic  potential model
 \cite{Dey1994}. Combining  with our previous work on the radiative decays $\Omega_Q^* \to \Omega_Q\, \gamma$ \cite{Wang0909},
 we perform systematic studies  for the
 radiative decays $B^*_6 \to B_{6}\,\gamma$ with the light-cone QCD sum rules. The strong decays $B^*_6 \to
B_{6}\,\pi$ are forbidden due to the unavailable phase space, while
the radiative channels are not phase space suppressed and become
relevant, although the electromagnetic strength is weaker than that
of the strong interaction. The properties of the charm baryon states
would be studied at the BESIII and $\rm{\bar{P}ANDA}$
\cite{BESIII,PANDA}, where the charm baryon states are copiously
produced at the $e^+e^-$ and $p\bar{p}$ collisions. The LHCb is a
dedicated $b$ and $c$-physics precision experiment at the LHC (large
hadron collider). The LHC will be the world's most copious  source
of the $b$ hadrons, and  a complete spectrum of the $b$ hadrons will
be available through gluon fusion. In proton-proton collisions at
$\sqrt{s}=14\,\rm{TeV}$, the $b\bar{b}$ cross section is expected to
be $\sim 500\mu b$ producing $10^{12}$ $b\bar{b}$ pairs in a
standard  year of running at the LHCb operational luminosity of
$2\times10^{32} \rm{cm}^{-2} \rm{sec}^{-1}$ \cite{LHC}. The present
predictions for the radiative decays can be tested at the BESIII,
$\rm{\bar{P}ANDA}$ and LHCb.

\section{Conclusion}

In this article, we parameterize the vertexes $B_Q^*B_Q V$ with
three tensor structures due to Lorentz invariance, study the
corresponding three strong coupling constants with the light-cone
QCD sum rules, then assume the vector meson dominance of the
intermediate $\phi(1020)$, $\rho_0(770)$ and $\omega(782)$ as the
contributions from the $J/\psi$ and $\Upsilon$ are negligible in the
heavy quark limit, and calculate the radiative decay widths
$\Gamma_{B_Q^*\to B_Q \gamma}$. The predictions can be tested by the
experimental data  at the BESIII, $\rm{\bar{P}ANDA}$ and LHCb in the
future. Although the values of the strong coupling constants $g_2$
are not stable enough with variations of the Borel parameter, the
Borel parameter dependence of the radiative decay widths is  very
weak, as the main contributions come from the strong coupling
constants $g_1$.  The heavy quark symmetry works rather well for the
strong coupling constants $g_1$ and $\rm{G3}$. The  strong coupling
constants in the vertexes $B_Q^*B_Q V$ are basic parameters in
describing the interactions among the heavy baryon states, once
reasonable values are obtained, we can use them to perform
phenomenological analysis.

\section*{Acknowledgment}
This  work is supported by National Natural Science Foundation,
Grant Number 10775051, and Program for New Century Excellent Talents
in University, Grant Number NCET-07-0282, and Project Supported by
Chinese Universities Scientific Fund.

\section*{Appendix}

\subsection*{Appendix\,\,A}
The  30 sum rules for the strong coupling constants $g_1$, $g_2$ and
$\rm{G}3$ in different channels,
\begin{eqnarray}
g^{\Xi^*_Q\Xi'_Q\phi}_1&=&\frac{1}{\lambda_{\Xi'_Q}\lambda_{\Xi_Q^*}
\left(M_{\Xi'_Q}+M_{\Xi_Q^*}\right)}\exp{\frac{M_{\Xi'_Q}^2+M_{\Xi_Q^*}^2-2u_0\bar{u}_0M_\phi^2}{2M^2}}\nonumber \\
&&\left\{-\frac{u_0f_\phi M_\phi
g_{\perp}^{(v)}(\bar{u}_0)}{8\pi^2}M^4E_1(x)\int_0^1 dt t(1-t) e^{-\frac{\widetilde{m}_Q^2}{M^2}}\right.\nonumber\\
 &&+ \frac{u_0m_Q^2f_\phi M_\phi
g_{\perp}^{(v)}(\bar{u}_0)}{144M^2}\langle\frac{\alpha_sGG}{\pi}\rangle\int_0^1 dt \frac{1-t}{t^2} e^{-\frac{\widetilde{m}_Q^2}{M^2}}\nonumber\\
&&-\frac{u_0\widetilde{f}_\phi M_\phi
}{32\pi^2}M^4E_1(x) \frac{d}{du_0}g_\perp^{(a)}(\bar{u}_0)\int_0^1 dt t(1-t) e^{-\frac{\widetilde{m}_Q^2}{M^2}}\nonumber\\
 &&+\frac{u_0m_Q^2\widetilde{f}_\phi M_\phi
}{576M^2}\langle \frac{\alpha_sGG}{\pi}\rangle \frac{d}{du_0}g_\perp^{(a)}(\bar{u}_0)\int_0^1 dt \frac{1-t}{t^2} e^{-\frac{\widetilde{m}_Q^2}{M^2}}\nonumber\\
&&-\frac{\widetilde{f}_\phi M_\phi g_\perp^{(a)}(\bar{u}_0)
}{16\pi^2}M^4E_1(x) \int_0^1 dt t  e^{-\frac{\widetilde{m}_Q^2}{M^2}}\nonumber\\
 &&+\frac{m_Q^2\widetilde{f}_\phi M_\phi
g_\perp^{(a)}(\bar{u}_0)}{288M^2}\langle \frac{\alpha_sGG}{\pi}\rangle  \int_0^1 dt \frac{1}{t^2} e^{-\frac{\widetilde{m}_Q^2}{M^2}}\nonumber\\
&&  +\frac{u_0f_\phi M_\phi^3 }{16\pi^2} M^2E_0(x)\int_0^1 dt
t\int_0^{u_0} d\alpha_{\bar{s}}
\int_{u_0-\alpha_{\bar{s}}}^{1-\alpha_{\bar{s}}}  d\alpha_g \frac{(1-2v)\mathcal{A}(\alpha_i)+\mathcal{V}(\alpha_i)}{\alpha_g}  e^{-\frac{\widetilde{m}_Q^2}{M^2}}   \nonumber\\
&& \left. +\frac{f_\phi M_\phi }{16\pi^2} M^4E_1(x)\int_0^1 dt t
\frac{d}{du_0}\int_0^{u_0} d\alpha_{\bar{s}}
\int_{u_0-\alpha_{\bar{s}}}^{1-\alpha_{\bar{s}}}  d\alpha_g
(1-v)\frac{\mathcal{A}(\alpha_i)+\mathcal{V}(\alpha_i)}{\alpha_g}
e^{-\frac{\widetilde{m}_Q^2}{M^2}}  \right\} \nonumber
 \end{eqnarray}
 \begin{eqnarray}
&+&\frac{1}{\lambda_{\Xi'_Q}\lambda_{\Xi_Q^*}
\left(M_{\Xi'_Q}+M_{\Xi_Q^*}\right)}\exp{\frac{M_{\Xi'_Q}^2+M_{\Xi_Q^*}^2-2m_Q^2-2u_0\bar{u}_0M_\phi^2}{2M^2}}\nonumber \\
&&\left\{  -\frac{\langle\bar{q}q\rangle f_\phi^{\perp} M_\phi^2
\widetilde{C}_{\perp}(\bar{u}_0)}{6}+ \frac{\langle\bar{q}q\rangle
f_\phi^{\perp}
\phi_{\perp}(\bar{u}_0)}{6} M^2E_0(x) \right.\nonumber\\
 &&- \frac{\langle \bar{q}g_s \sigma
Gq\rangle
f_\phi^{\perp}\phi_{\perp}(\bar{u}_0)}{24}\left(1+\frac{m_Q^2}{M^2}\right)+
\frac{m_Q^4 f_\phi^{\perp} M_\phi^2 \langle \bar{q}g_s\sigma G
q\rangle
A_{\perp}(\bar{u}_0)}{96M^6}\nonumber\\
 &&\left.- \frac{\langle\bar{q}q\rangle f_\phi^{\perp} M_\phi^2
A_{\perp}(\bar{u}_0)}{24} \left(1+\frac{m_Q^2}{M^2}\right)
+\frac{u_0\langle\bar{q}g_s \sigma Gq\rangle f_\phi^{\perp} M_\phi^2
 \widetilde{C}_{\perp}(\bar{u}_0)}{24M^2} \left(1+\frac{m_Q^2}{M^2}\right) \right\} \,
 , \nonumber\\
 \end{eqnarray}

\begin{eqnarray}
g^{\Xi_Q^*\Xi'_Q\phi}_2&=&\frac{1}{\lambda_{\Xi'_Q}\lambda_{\Xi_Q^*}}\exp{\frac{M_{\Xi'_Q}^2+M_{\Xi_Q^*}^2-2u_0\bar{u}_0M_\phi^2}{2M^2}}\nonumber \\
&&\left\{\frac{u_0f_\phi M_\phi
 \left[\widetilde{\phi}_{\parallel}(\bar{u}_0)-\widetilde{g}_{\perp}^{(v)}(\bar{u}_0)\right]}{4\pi^2}M^2E_0(x)\int_0^1 dt t(1-t) e^{-\frac{\widetilde{m}_Q^2}{M^2}}\right.\nonumber\\
 &&- \frac{u_0m_Q^2 f_\phi M_\phi
\left[\widetilde{\phi}_{\parallel}(\bar{u}_0)-\widetilde{g}_{\perp}^{(v)}(\bar{u}_0)\right]}{72M^4}\langle\frac{\alpha_sGG}{\pi}\rangle\int_0^1 dt \frac{1-t}{t^2} e^{-\frac{\widetilde{m}_Q^2}{M^2}}\nonumber\\
&& -\frac{u_0f_\phi M_\phi^3 \widetilde{A}(\bar{u}_0)}{16\pi^2}
\int_0^1 dt t e^{-\frac{\widetilde{m}_Q^2}{M^2}}
+\frac{u_0m_Q^2f_\phi M_\phi^3 \widetilde{A}(\bar{u}_0)}{288M^6}
\langle \frac{\alpha_sGG}{\pi}
\rangle\int_0^1 dt \frac{1}{t^2} e^{-\frac{\widetilde{m}_Q^2}{M^2}} \nonumber\\
 &&-\frac{u_0\widetilde{f}_\phi M_\phi
}{16\pi^2}M^2E_0(x) g_\perp^{(a)}(\bar{u}_0)\int_0^1 dt t(1-t) e^{-\frac{\widetilde{m}_Q^2}{M^2}}\nonumber\\
 &&+\frac{u_0m_Q^2\widetilde{f}_\phi M_\phi
}{288M^4}\langle \frac{\alpha_sGG}{\pi}\rangle g_\perp^{(a)}(\bar{u}_0)\int_0^1 dt \frac{1-t}{t^2} e^{-\frac{\widetilde{m}_Q^2}{M^2}}\nonumber\\
&&  \left.+\frac{f_\phi M_\phi}{8\pi^2} M^2E_0(x) \int_0^1 dt
t\int_0^{u_0} d\alpha_{\bar{s}}
\int_{u_0-\alpha_{\bar{s}}}^{1-\alpha_{\bar{s}}} d\alpha_g
\frac{\mathcal{A}(\alpha_i)+(1-2v)\mathcal{V}(\alpha_i)} {\alpha_g}
e^{-\frac{\widetilde{m}_Q^2}{M^2}}    \right\} \nonumber
 \end{eqnarray}
 \begin{eqnarray}
&+&\frac{1}{\lambda_{\Xi'_Q}\lambda_{\Xi_Q^*}
}\exp{\frac{M_{\Xi'_Q}^2+M_{\Xi_Q^*}^2-2m_Q^2-2u_0\bar{u}_0 M_\phi^2}{2M^2}}\nonumber \\
&&\left\{\frac{2u_0\langle\bar{q}q\rangle f_\phi^{\perp}M_\phi^2
\widetilde{\widetilde{B}}_{\perp}(\bar{u}_0) }{3M^2} -
\frac{u_0\langle \bar{q}g_s \sigma Gq\rangle
f_\phi^{\perp}M_\phi^2\widetilde{\widetilde{B}}_{\perp}(\bar{u}_0)}{6M^4}\left(1+\frac{m_Q^2}{M^2}\right)\right\}
 \, ,
 \end{eqnarray}

\begin{eqnarray}
\rm{G}3_{\Xi_Q^*\Xi'_Q\phi}&=&\frac{1}{\lambda_{\Xi'_Q}\lambda_{\Xi_Q^*}}\exp{\frac{M_{\Xi'_Q}^2+M_{\Xi_Q^*}^2-2u_0
\bar{u}_0 M_\phi^2}{2M^2}}\nonumber \\
&&\left\{\frac{f_\phi M_\phi
 \left[\widetilde{\phi}_{\parallel}(\bar{u}_0)-\widetilde{g}_{\perp}^{(v)}(\bar{u}_0)\right]}{8\pi^2}M^4E_1(x)\int_0^1 dt t(1-t) e^{-\frac{\widetilde{m}_Q^2}{M^2}}\right.\nonumber\\
 &&- \frac{m_Q^2f_\phi M_\phi
\left[\widetilde{\phi}_{\parallel}(\bar{u}_0)-\widetilde{g}_{\perp}^{(v)}(\bar{u}_0)\right]}{144M^2}\langle\frac{\alpha_sGG}{\pi}\rangle\int_0^1 dt \frac{1-t}{t^2} e^{-\frac{\widetilde{m}_Q^2}{M^2}}\nonumber\\
&& -\frac{f_\phi M_\phi^3  \widetilde{A}(\bar{u}_0)}{32\pi^2}
M^2E_0(x)\int_0^1 dt t
e^{-\frac{\widetilde{m}_Q^2}{M^2}}\nonumber\\
 &&+\frac{m_Q^2f_\phi M_\phi^3 \widetilde{A}(\bar{u}_0)}{576M^4}
\langle \frac{\alpha_sGG}{\pi}
\rangle\int_0^1 dt \frac{1}{t^2} e^{-\frac{\widetilde{m}_Q^2}{M^2}} \nonumber\\
&&+\frac{\widetilde{f}_\phi M_\phi
}{32\pi^2}M^4E_1(x) g_\perp^{(a)}(\bar{u}_0)\int_0^1 dt t(1+t) e^{-\frac{\widetilde{m}_Q^2}{M^2}}\nonumber\\
 &&-\frac{m_Q^2\widetilde{f}_\phi M_\phi
}{576M^2}\langle \frac{\alpha_sGG}{\pi}\rangle g_\perp^{(a)}(\bar{u}_0)\int_0^1 dt \frac{1+t}{t^2} e^{-\frac{\widetilde{m}_Q^2}{M^2}}\nonumber\\
&&  \left.-\frac{u_0f_\phi M_\phi^3 }{8\pi^2}M^2E_0(x) \int_0^1 dt
t\int_0^{u_0} d\alpha_{\bar{s}}
\int_{u_0-\alpha_{\bar{s}}}^{1-\alpha_{\bar{s}}} d\alpha_g
v\frac{\mathcal{A}(\alpha_i)-\mathcal{V}(\alpha_i)} {\alpha_g}
e^{-\frac{\widetilde{m}_Q^2}{M^2}}    \right\} \nonumber
 \end{eqnarray}
\begin{eqnarray}
&+&\frac{1}{\lambda_{\Xi'_Q}\lambda_{\Xi_Q^*}
}\exp{\frac{M_{\Xi'_Q}^2+M_{\Xi_Q^*}^2-2m_Q^2-2u_0\bar{u}_0M_\phi^2}{2M^2}}\nonumber \\
&&\left\{-\frac{\langle\bar{q}q\rangle f_\phi^{\perp}
\phi_{\perp}(\bar{u}_0)}{6}M^2E_0(x) + \frac{\langle \bar{q}g_s
\sigma
Gq\rangle f_\phi^{\perp}\phi_{\perp}(\bar{u}_0)}{24}\left(1+\frac{m_Q^2}{M^2}\right) \right.\nonumber\\
 &&- \frac{m_Q^4 f_\phi^{\perp} M_\phi^2 \langle \bar{q}g_s\sigma G q\rangle
A_{\perp}(\bar{u}_0)}{96M^6}  + \frac{\langle\bar{q}q\rangle
f_\phi^{\perp} M_\phi^2
A_{\perp}(\bar{u}_0)}{24} \left(1+\frac{m_Q^2}{M^2}\right)   \nonumber\\
 &&\left. +\frac{\langle\bar{q}q\rangle f_\phi^{\perp}M_\phi^2
\widetilde{\widetilde{B}}_{\perp}(\bar{u}_0) }{3}- \frac{\langle
\bar{q}g_s \sigma Gq\rangle
f_\phi^{\perp}M_\phi^2\widetilde{\widetilde{B}}_{\perp}(\bar{u}_0)}{12M^2}\left(1+\frac{m_Q^2}{M^2}\right)
  \right\} \, ,
 \end{eqnarray}

\begin{eqnarray}
g^{\Xi_Q^*\Xi'_Q\rho_0}_1&=&\pm\frac{1}{\lambda_{\Xi'_Q}\lambda_{\Xi_Q^*}
\left(M_{\Xi'_Q}+M_{\Xi_Q^*}\right)}\exp{\frac{M_{\Xi'_Q}^2+M_{\Xi_Q^*}^2-2u_0\bar{u}_0M_\rho^2}{2M^2}}\nonumber \\
&&\left\{-\frac{u_0f_\rho M_\rho
g_{\perp}^{(v)}(\bar{u}_0)}{8\pi^2}M^4E_1(x)\int_0^1 dt t(1-t) e^{-\frac{\widetilde{m}_Q^2}{M^2}}\right.\nonumber\\
 &&+ \frac{u_0m_Q^2f_\rho M_\rho
g_{\perp}^{(v)}(\bar{u}_0)}{144M^2}\langle\frac{\alpha_sGG}{\pi}\rangle\int_0^1 dt \frac{1-t}{t^2} e^{-\frac{\widetilde{m}_Q^2}{M^2}}\nonumber\\
&&-\frac{m_sf_\rho^{\perp}
\phi_{\perp}(\bar{u}_0)}{8\pi^2}M^4E_1(x)\int_0^1 dt t e^{-\frac{\widetilde{m}_Q^2}{M^2}}\nonumber\\
 &&+ \frac{m_sm_Q^2f_\rho^{\perp}
\phi_{\perp}(\bar{u}_0)}{144M^2}\langle\frac{\alpha_sGG}{\pi}\rangle\int_0^1
dt \frac{1}{t^2} e^{-\frac{\widetilde{m}_Q^2}{M^2}} \nonumber \\
&& +\frac{u_0m_sf_\rho^{\perp}M_\rho^2
\widetilde{C}_{\perp}(\bar{u}_0)}{8\pi^2}M^2E_0(x)\int_0^1 dt t e^{-\frac{\widetilde{m}_Q^2}{M^2}}\nonumber\\
 &&- \frac{u_0m_sm_Q^2f_\rho^{\perp}M_\rho^2
\widetilde{C}_{\perp}(\bar{u}_0)}{144M^4}\langle\frac{\alpha_sGG}{\pi}\rangle\int_0^1 dt \frac{1}{t^2} e^{-\frac{\widetilde{m}_Q^2}{M^2}}\nonumber\\
&&-\frac{u_0\widetilde{f}_\rho M_\rho
}{32\pi^2}M^4E_1(x) \frac{d}{du_0}g_\perp^{(a)}(\bar{u}_0)\int_0^1 dt t(1-t) e^{-\frac{\widetilde{m}_Q^2}{M^2}}\nonumber\\
 &&+\frac{u_0m_Q^2\widetilde{f}_\rho M_\rho
}{576M^2}\langle \frac{\alpha_sGG}{\pi}\rangle \frac{d}{du_0}g_\perp^{(a)}(\bar{u}_0)\int_0^1 dt \frac{1-t}{t^2} e^{-\frac{\widetilde{m}_Q^2}{M^2}}\nonumber\\
&&-\frac{\widetilde{f}_\rho M_\rho g_\perp^{(a)}(\bar{u}_0)
}{16\pi^2}M^4E_1(x) \int_0^1 dt t  e^{-\frac{\widetilde{m}_Q^2}{M^2}}\nonumber\\
 &&+\frac{m_Q^2\widetilde{f}_\rho M_\rho
g_\perp^{(a)}(\bar{u}_0)}{144M^2}\langle \frac{\alpha_sGG}{\pi}\rangle  \int_0^1 dt \frac{1}{t^2} e^{-\frac{\widetilde{m}_Q^2}{M^2}}\nonumber\\
&&  +\frac{u_0f_\rho M_\rho^3 }{16\pi^2} M^2E_0(x)\int_0^1 dt
t\int_0^{u_0} d\alpha_{\bar{u}}
\int_{u_0-\alpha_{\bar{u}}}^{1-\alpha_{\bar{u}}}  d\alpha_g \frac{(1-2v)\mathcal{A}(\alpha_i)+\mathcal{V}(\alpha_i)}{\alpha_g}  e^{-\frac{\widetilde{m}_Q^2}{M^2}}   \nonumber\\
&& \left. +\frac{f_\rho M_\rho }{16\pi^2} M^4E_1(x)\int_0^1 dt t
\frac{d}{du_0}\int_0^{u_0} d\alpha_{\bar{u}}
\int_{u_0-\alpha_{\bar{u}}}^{1-\alpha_{\bar{u}}}  d\alpha_g
(1-v)\frac{\mathcal{A}(\alpha_i)+\mathcal{V}(\alpha_i)}{\alpha_g}
e^{-\frac{\widetilde{m}_Q^2}{M^2}}  \right\} \nonumber
 \end{eqnarray}
 \begin{eqnarray}
&\pm&\frac{1}{\lambda_{\Xi'_Q}\lambda_{\Xi_Q^*}
\left(M_{\Xi'_Q}+M_{\Xi_Q^*}\right)}\exp{\frac{M_{\Xi'_Q}^2+M_{\Xi_Q^*}^2-2m_Q^2-2u_0\bar{u}_0M_\rho^2}{2M^2}}\nonumber \\
&&\left\{-\frac{u_0m_s\langle\bar{s}s\rangle f_\rho M_\rho
g_{\perp}^{(v)}(\bar{u}_0)}{12}  -\frac{\langle\bar{s}s\rangle
f_\rho^{\perp} M_\rho^2
\widetilde{C}_{\perp}(\bar{u}_0)}{6}  \right.\nonumber\\
&&+ \frac{u_0m_s\langle \bar{s}g_s \sigma Gs\rangle f_\rho M_\rho
g_{\perp}^{(v)}(\bar{u}_0)}{72M^2}\left(1+\frac{m_Q^2}{M^2}\right)
\nonumber\\
 &&+\frac{\langle\bar{s}s\rangle f_\rho^{\perp}
\phi_{\perp}(\bar{u}_0)}{6} M^2E_0(x)- \frac{\langle \bar{s}g_s
\sigma
Gs\rangle f_\rho^{\perp}\phi_{\perp}(\bar{u}_0)}{24}\left(1+\frac{m_Q^2}{M^2}\right)\nonumber\\
 &&+ \frac{m_Q^4 f_\rho^{\perp} M_\rho^2 \langle \bar{s}g_s\sigma G s\rangle
A_{\perp}(\bar{u}_0)}{96M^6} + \frac{m_s f_\rho^{\perp} M_\rho^2
A_{\perp}(\bar{u}_0)}{32\pi^2} M^2 E_0(x)   \nonumber\\
&&- \frac{\langle\bar{s}s\rangle f_\rho^{\perp} M_\rho^2
A_{\perp}(\bar{u}_0)}{24} \left(1+\frac{m_Q^2}{M^2}\right)
+\frac{u_0\langle\bar{s}g_s \sigma Gs\rangle f_\rho^{\perp} M_\rho^2
 \widetilde{C}_{\perp}(\bar{u}_0)}{24M^2} \left(1+\frac{m_Q^2}{M^2}\right)\nonumber\\
&&\left.-\frac{m_s\langle \bar{s}s\rangle\widetilde{f}_\rho M_\rho
g_\perp^{(a)}(\bar{u}_0)}{24} \left(1+\frac{m_Q^2}{M^2}\right)
-\frac{u_0m_s\langle \bar{s}s\rangle\widetilde{f}_\rho M_\rho }{48}
\frac{d}{du_0}g_\perp^{(a)}(\bar{u}_0) \right\} \, ,
 \end{eqnarray}

 \begin{eqnarray}
g^{\Xi_Q^*\Xi'_Q\rho_0}_2&=\pm&\frac{1}{\lambda_{\Xi'_Q}\lambda_{\Xi_Q^*}}\exp{\frac{M_{\Xi'_Q}^2+M_{\Xi_Q^*}^2
-2u_0\bar{u}_0M_\rho^2}{2M^2}}\nonumber \\
&&\left\{\frac{u_0f_\rho M_\rho
 \left[\widetilde{\phi}_{\parallel}(\bar{u}_0)-\widetilde{g}_{\perp}^{(v)}(\bar{u}_0)\right]}{4\pi^2}M^2E_0(x)\int_0^1 dt t(1-t) e^{-\frac{\widetilde{m}_Q^2}{M^2}}\right.\nonumber\\
 &&- \frac{u_0m_Q^2 f_\rho M_\rho
\left[\widetilde{\phi}_{\parallel}(\bar{u}_0)-\widetilde{g}_{\perp}^{(v)}(\bar{u}_0)\right]}{72M^4}\langle\frac{\alpha_sGG}{\pi}\rangle\int_0^1 dt \frac{1-t}{t^2} e^{-\frac{\widetilde{m}_Q^2}{M^2}}\nonumber\\
&& -\frac{u_0f_\rho M_\rho^3 \widetilde{A}(\bar{u}_0)}{16\pi^2}
\int_0^1 dt t e^{-\frac{\widetilde{m}_Q^2}{M^2}}
+\frac{u_0m_Q^2f_\rho M_\rho^3 \widetilde{A}(\bar{u}_0)}{288M^6}
\langle \frac{\alpha_sGG}{\pi}
\rangle\int_0^1 dt \frac{1}{t^2} e^{-\frac{\widetilde{m}_Q^2}{M^2}} \nonumber\\
 &&-\frac{u_0m_sf_\rho^{\perp}M_\rho^2
\widetilde{\widetilde{B}}_{\perp}(\bar{u}_0)}{2\pi^2}\int_0^1 dt t e^{-\frac{\widetilde{m}_Q^2}{M^2}}\nonumber\\
 &&+ \frac{u_0m_sm_Q^2f_\rho^{\perp}M_\rho^2
\widetilde{\widetilde{B}}_{\perp}(\bar{u}_0)}{36M^6}\langle\frac{\alpha_sGG}{\pi}\rangle\int_0^1 dt \frac{1}{t^2} e^{-\frac{\widetilde{m}_Q^2}{M^2}}\nonumber\\
&&-\frac{u_0\widetilde{f}_\rho M_\rho
}{16\pi^2}M^2E_0(x) g_\perp^{(a)}(\bar{u}_0)\int_0^1 dt t(1-t) e^{-\frac{\widetilde{m}_Q^2}{M^2}}\nonumber\\
 &&+\frac{u_0m_Q^2\widetilde{f}_\rho M_\rho
}{288M^4}\langle \frac{\alpha_sGG}{\pi}\rangle g_\perp^{(a)}(\bar{u}_0)\int_0^1 dt \frac{1-t}{t^2} e^{-\frac{\widetilde{m}_Q^2}{M^2}}\nonumber\\
&&  \left.+\frac{f_\rho M_\rho}{8\pi^2} M^2E_0(x) \int_0^1 dt
t\int_0^{u_0} d\alpha_{\bar{u}}
\int_{u_0-\alpha_{\bar{u}}}^{1-\alpha_{\bar{u}}} d\alpha_g
\frac{\mathcal{A}(\alpha_i)+(1-2v)\mathcal{V}(\alpha_i)} {\alpha_g}
e^{-\frac{\widetilde{m}_Q^2}{M^2}}    \right\} \nonumber
 \end{eqnarray}
 \begin{eqnarray}
&\pm&\frac{1}{\lambda_{\Xi'_Q}\lambda_{\Xi_Q^*}
}\exp{\frac{M_{\Xi'_Q}^2+M_{\Xi_Q^*}^2-2m_Q^2-2u_0\bar{u}_0M_\rho^2}{2M^2}}\nonumber \\
&&\left\{\frac{u_0m_s\langle\bar{s}s\rangle f_\rho M_\rho
\left[\widetilde{\phi}_\parallel(\bar{u}_0)-\widetilde{g}_{\perp}^{(v)}(\bar{u}_0)\right]}{6M^2}
  \right.\nonumber\\
&&- \frac{u_0m_s\langle \bar{s} s\rangle f_\rho M_\rho^3
\widetilde{A}(\bar{u}_0)}{24M^4}\left(1+\frac{m_Q^2}{M^2}\right)
\nonumber\\
&&- \frac{u_0m_s\langle \bar{s}g_s \sigma Gs\rangle f_\rho M_\rho
\left[\widetilde{\phi}_\parallel(\bar{u}_0)-\widetilde{g}_{\perp}^{(v)}(\bar{u}_0)\right]}{36M^4}\left(1+\frac{m_Q^2}{M^2}\right)
\nonumber\\
 &&+\frac{2u_0\langle\bar{s}s\rangle f_\rho^{\perp}M_\rho^2
\widetilde{\widetilde{B}}_{\perp}(\bar{u}_0) }{3M^2} -
\frac{u_0\langle \bar{s}g_s \sigma
Gs\rangle f_\rho^{\perp}M_\rho^2\widetilde{\widetilde{B}}_{\perp}(\bar{u}_0)}{6M^4}\left(1+\frac{m_Q^2}{M^2}\right)\nonumber\\
 &&\left.-\frac{u_0m_s\langle
\bar{s}s\rangle\widetilde{f}_\rho M_\rho
g_\perp^{(a)}(\bar{u}_0)}{24M^2} \right\} \, ,
 \end{eqnarray}

\begin{eqnarray}
\rm{G}3_{\Xi^*_Q\Xi'_Q\rho_0}&=&\pm\frac{1}{\lambda_{\Xi'_Q}\lambda_{\Xi_Q^*}}\exp{\frac{M_{\Xi'_Q}^2+M_{\Xi_Q^*}^2
-2u_0\bar{u}_0M_\rho^2}{2M^2}}\nonumber \\
&&\left\{\frac{f_\rho M_\rho
 \left[\widetilde{\phi}_{\parallel}(\bar{u}_0)-\widetilde{g}_{\perp}^{(v)}(\bar{u}_0)\right]}{8\pi^2}M^4E_1(x)\int_0^1 dt t(1-t) e^{-\frac{\widetilde{m}_Q^2}{M^2}}\right.\nonumber\\
 &&- \frac{m_Q^2f_\rho M_\rho
\left[\widetilde{\phi}_{\parallel}(\bar{u}_0)-\widetilde{g}_{\perp}^{(v)}(\bar{u}_0)\right]}{144M^2}\langle\frac{\alpha_sGG}{\pi}\rangle\int_0^1 dt \frac{1-t}{t^2} e^{-\frac{\widetilde{m}_Q^2}{M^2}}\nonumber\\
&& -\frac{f_\rho M_\rho^3  \widetilde{A}(\bar{u}_0)}{32\pi^2}
M^2E_0(x)\int_0^1 dt t
e^{-\frac{\widetilde{m}_Q^2}{M^2}}\nonumber\\
 &&+\frac{m_Q^2f_\rho M_\rho^3 \widetilde{A}(\bar{u}_0)}{576M^4}
\langle \frac{\alpha_sGG}{\pi}
\rangle\int_0^1 dt \frac{1}{t^2} e^{-\frac{\widetilde{m}_Q^2}{M^2}} \nonumber\\
&&+\frac{m_sf_\rho^{\perp}
\phi_{\perp}(\bar{u}_0)}{8\pi^2}M^4E_1(x)\int_0^1 dt t
e^{-\frac{\widetilde{m}_Q^2}{M^2}}\nonumber\\
 &&- \frac{m_sm_Q^2f_\rho^{\perp}
\phi_{\perp}(\bar{u}_0)}{144M^2}\langle\frac{\alpha_sGG}{\pi}\rangle\int_0^1 dt \frac{1}{t^2} e^{-\frac{\widetilde{m}_Q^2}{M^2}}\nonumber\\
 &&-\frac{m_sf_\rho^{\perp}M_\rho^2\widetilde{\widetilde{B}}_\perp
 (\bar{u}_0)}{4\pi^2}M^2E_0(x)\int_0^1 dt t e^{-\frac{\widetilde{m}_Q^2}{M^2}}\nonumber\\
 &&+ \frac{m_sm_Q^2f_\rho^{\perp}M_\rho^2
\widetilde{\widetilde{B}}_{\perp}(\bar{u}_0)}{72M^4}\langle\frac{\alpha_sGG}{\pi}\rangle\int_0^1 dt \frac{1}{t^2} e^{-\frac{\widetilde{m}_Q^2}{M^2}}\nonumber\\
&&+\frac{\widetilde{f}_\rho M_\rho
}{32\pi^2}M^4E_1(x) g_\perp^{(a)}(\bar{u}_0)\int_0^1 dt t(1+t) e^{-\frac{\widetilde{m}_Q^2}{M^2}}\nonumber\\
 &&-\frac{m_Q^2\widetilde{f}_\rho M_\rho
}{576M^2}\langle \frac{\alpha_sGG}{\pi}\rangle g_\perp^{(a)}(\bar{u}_0)\int_0^1 dt \frac{1+t}{t^2} e^{-\frac{\widetilde{m}_Q^2}{M^2}}\nonumber\\
&&  \left.-\frac{u_0f_\rho M_\rho^3 }{8\pi^2}M^2E_0(x) \int_0^1 dt
t\int_0^{u_0} d\alpha_{\bar{u}}
\int_{u_0-\alpha_{\bar{u}}}^{1-\alpha_{\bar{u}}} d\alpha_g
v\frac{\mathcal{A}(\alpha_i)-\mathcal{V}(\alpha_i)} {\alpha_g}
e^{-\frac{\widetilde{m}_Q^2}{M^2}}    \right\} \nonumber
 \end{eqnarray}
\begin{eqnarray}
&\pm&\frac{1}{\lambda_{\Xi'_Q}\lambda_{\Xi_Q^*}
}\exp{\frac{M_{\Xi'_Q}^2+M_{\Xi_Q^*}^2-2m_Q^2-2u_0\bar{u}_0M_\rho^2}{2M^2}}\nonumber \\
&&\left\{\frac{m_s\langle\bar{s}s\rangle f_\rho M_\rho
\left[\widetilde{\phi}_\parallel(\bar{u}_0)-\widetilde{g}_{\perp}^{(v)}(\bar{u}_0)\right]}{12}
  \right.\nonumber\\
&&- \frac{m_s\langle \bar{s}s\rangle f_\rho M_\rho^3
\widetilde{A}(\bar{u}_0)}{48M^2}\left(1+\frac{m_Q^2}{M^2}\right)
\nonumber\\
&&- \frac{m_s\langle \bar{s}g_s \sigma Gs\rangle f_\rho M_\rho
\left[\widetilde{\phi}_\parallel(\bar{u}_0)-\widetilde{g}_{\perp}^{(v)}(\bar{u}_0)\right]}{72M^2}\left(1+\frac{m_Q^2}{M^2}\right)
\nonumber\\
 &&-\frac{\langle\bar{s}s\rangle f_\rho^{\perp}
\phi_{\perp}(\bar{u}_0)}{6}M^2E_0(x) + \frac{\langle \bar{s}g_s
\sigma
Gs\rangle f_\rho^{\perp}\phi_{\perp}(\bar{u}_0)}{24}\left(1+\frac{m_Q^2}{M^2}\right)\nonumber\\
 &&- \frac{m_Q^4 f_\rho^{\perp} M_\rho^2 \langle \bar{s}g_s\sigma G s\rangle
A_{\perp}(\bar{u}_0)}{96M^6} - \frac{m_s f_\rho^{\perp} M_\rho^2
A_{\perp}(\bar{u}_0)}{32\pi^2} M^2 E_0(x)   \nonumber\\
&&+ \frac{\langle\bar{s}s\rangle f_\rho^{\perp} M_\rho^2
A_{\perp}(\bar{u}_0)}{24} \left(1+\frac{m_Q^2}{M^2}\right)
+\frac{\langle\bar{s}s\rangle f_\rho^{\perp}M_\rho^2
\widetilde{\widetilde{B}}_{\perp}(\bar{u}_0) }{3}\nonumber\\
 && - \frac{\langle
\bar{s}g_s \sigma
Gs\rangle f_\rho^{\perp}M_\rho^2\widetilde{\widetilde{B}}_{\perp}(\bar{u}_0)}{12M^2}\left(1+\frac{m_Q^2}{M^2}\right)\nonumber\\
 &&\left.+\frac{m_s\langle
\bar{s}s\rangle\widetilde{f}_\rho M_\rho
g_\perp^{(a)}(\bar{u}_0)}{48}\left(1+\frac{2m_Q^2}{M^2}\right)
\right\} \, ,
 \end{eqnarray}

 \begin{eqnarray}
g^{\Sigma_Q^*\Sigma_Q\rho_0}_1&=&\pm\frac{1}{\lambda_{\Sigma_Q}\lambda_{\Sigma_Q^*}
\left(M_{\Sigma_Q}+M_{\Sigma_Q^*}\right)}\exp{\frac{M_{\Sigma_Q}^2+M_{\Sigma_Q^*}^2-2u_0 \bar{u}_0 M_\rho^2}{2M^2}}\nonumber \\
&&\left\{-\frac{u_0f_\rho M_\rho
g_{\perp}^{(v)}(\bar{u}_0)}{2\pi^2}M^4E_1(x)\int_0^1 dt t(1-t) e^{-\frac{\widetilde{m}_Q^2}{M^2}}\right.\nonumber\\
 &&+ \frac{u_0m_Q^2f_\rho M_\rho
g_{\perp}^{(v)}(\bar{u}_0)}{36M^2}\langle\frac{\alpha_sGG}{\pi}\rangle\int_0^1 dt \frac{1-t}{t^2} e^{-\frac{\widetilde{m}_Q^2}{M^2}}\nonumber\\
&&-\frac{u_0\widetilde{f}_\rho M_\rho
}{8\pi^2}M^4E_1(x) \frac{d}{du_0}g_\perp^{(a)}(\bar{u}_0)\int_0^1 dt t(1-t) e^{-\frac{\widetilde{m}_Q^2}{M^2}}\nonumber\\
 &&+\frac{u_0m_Q^2\widetilde{f}_\rho M_\rho
}{144M^2}\langle \frac{\alpha_sGG}{\pi}\rangle \frac{d}{du_0}g_\perp^{(a)}(\bar{u}_0)\int_0^1 dt \frac{1-t}{t^2} e^{-\frac{\widetilde{m}_Q^2}{M^2}}\nonumber\\
&&-\frac{\widetilde{f}_\rho M_\rho g_\perp^{(a)}(\bar{u}_0)
}{4\pi^2}M^4E_1(x) \int_0^1 dt t  e^{-\frac{\widetilde{m}_Q^2}{M^2}}\nonumber\\
 &&+\frac{m_Q^2\widetilde{f}_\rho M_\rho
g_\perp^{(a)}(\bar{u}_0)}{72M^2}\langle \frac{\alpha_sGG}{\pi}\rangle  \int_0^1 dt \frac{1}{t^2} e^{-\frac{\widetilde{m}_Q^2}{M^2}}\nonumber\\
&&  +\frac{u_0f_\rho M_\rho^3 }{4\pi^2} M^2E_0(x)\int_0^1 dt
t\int_0^{u_0} d\alpha_{\bar{u}}
\int_{u_0-\alpha_{\bar{u}}}^{1-\alpha_{\bar{u}}}  d\alpha_g \frac{(1-2v)\mathcal{A}(\alpha_i)+\mathcal{V}(\alpha_i)}{\alpha_g}  e^{-\frac{\widetilde{m}_Q^2}{M^2}}   \nonumber\\
&& \left. +\frac{f_\rho M_\rho }{4\pi^2} M^4E_1(x)\int_0^1 dt t
\frac{d}{du_0}\int_0^{u_0} d\alpha_{\bar{u}}
\int_{u_0-\alpha_{\bar{u}}}^{1-\alpha_{\bar{u}}}  d\alpha_g
(1-v)\frac{\mathcal{A}(\alpha_i)+\mathcal{V}(\alpha_i)}{\alpha_g}
e^{-\frac{\widetilde{m}_Q^2}{M^2}}  \right\} \nonumber
 \end{eqnarray}
 \begin{eqnarray}
&\pm&\frac{1}{\lambda_{\Sigma_Q}\lambda_{\Sigma_Q^*}
\left(M_{\Sigma_Q}+M_{\Sigma_Q^*}\right)}\exp{\frac{M_{\Sigma_Q}^2+M_{\Sigma_Q^*}^2-2m_Q^2-2u_0 \bar{u}_0 M_\rho^2}{2M^2}}\nonumber \\
&&\left\{   -\frac{2\langle\bar{q}q\rangle f_\rho^{\perp} M_\rho^2
\widetilde{C}_{\perp}(\bar{u}_0)}{3}+\frac{2\langle\bar{q}q\rangle
f_\rho^{\perp}
\phi_{\perp}(\bar{u}_0)}{3} M^2E_0(x)  \right.\nonumber\\
  &&- \frac{\langle \bar{q}g_s \sigma
Gq\rangle
f_\rho^{\perp}\phi_{\perp}(\bar{u}_0)}{6}\left(1+\frac{m_Q^2}{M^2}\right)+\frac{m_Q^4
f_\rho^{\perp} M_\rho^2 \langle \bar{q}g_s\sigma G q\rangle
A_{\perp}(\bar{u}_0)}{24M^6}\nonumber\\
 &&\left.- \frac{\langle\bar{q}q\rangle f_\rho^{\perp} M_\rho^2
A_{\perp}(\bar{u}_0)}{6} \left(1+\frac{m_Q^2}{M^2}\right)
+\frac{u_0\langle\bar{q}g_s \sigma Gq\rangle f_\rho^{\perp} M_\rho^2
 \widetilde{C}_{\perp}(\bar{u}_0)}{6M^2} \left(1+\frac{m_Q^2}{M^2}\right) \right\} \,
 , \nonumber\\
 \end{eqnarray}

 \begin{eqnarray}
g^{\Sigma_Q^*\Sigma_Q\rho_0}_2&=&\pm\frac{1}{\lambda_{\Sigma_Q}\lambda_{\Sigma_Q^*}}\exp{\frac{M_{\Sigma_Q}^2+
M_{\Sigma_Q^*}^2-2u_0\bar{u}_0M_\rho^2}{2M^2}}\nonumber \\
&&\left\{\frac{u_0f_\rho M_\rho
 \left[\widetilde{\phi}_{\parallel}(\bar{u}_0)-\widetilde{g}_{\perp}^{(v)}(\bar{u}_0)\right]}{\pi^2}M^2E_0(x)\int_0^1 dt t(1-t) e^{-\frac{\widetilde{m}_Q^2}{M^2}}\right.\nonumber\\
 &&- \frac{u_0m_Q^2 f_\rho M_\rho
\left[\widetilde{\phi}_{\parallel}(\bar{u}_0)-\widetilde{g}_{\perp}^{(v)}(\bar{u}_0)\right]}{18M^4}\langle\frac{\alpha_sGG}{\pi}\rangle\int_0^1 dt \frac{1-t}{t^2} e^{-\frac{\widetilde{m}_Q^2}{M^2}}\nonumber\\
&& -\frac{u_0f_\rho M_\rho^3 \widetilde{A}(\bar{u}_0)}{4\pi^2}
\int_0^1 dt t e^{-\frac{\widetilde{m}_Q^2}{M^2}}
+\frac{u_0m_Q^2f_\rho M_\rho^3 \widetilde{A}(\bar{u}_0)}{72M^6}
\langle \frac{\alpha_sGG}{\pi}
\rangle\int_0^1 dt \frac{1}{t^2} e^{-\frac{\widetilde{m}_Q^2}{M^2}} \nonumber\\
&&-\frac{u_0\widetilde{f}_\rho M_\rho
}{4\pi^2}M^2E_0(x) g_\perp^{(a)}(\bar{u}_0)\int_0^1 dt t(1-t) e^{-\frac{\widetilde{m}_Q^2}{M^2}}\nonumber\\
 &&+\frac{u_0m_Q^2\widetilde{f}_\rho M_\rho
}{72M^4}\langle \frac{\alpha_sGG}{\pi}\rangle g_\perp^{(a)}(\bar{u}_0)\int_0^1 dt \frac{1-t}{t^2} e^{-\frac{\widetilde{m}_Q^2}{M^2}}\nonumber\\
&&  \left.+\frac{f_\rho M_\rho}{2\pi^2} M^2E_0(x) \int_0^1 dt
t\int_0^{u_0} d\alpha_{\bar{u}}
\int_{u_0-\alpha_{\bar{u}}}^{1-\alpha_{\bar{u}}} d\alpha_g
\frac{\mathcal{A}(\alpha_i)+(1-2v)\mathcal{V}(\alpha_i)} {\alpha_g}
e^{-\frac{\widetilde{m}_Q^2}{M^2}}    \right\} \nonumber
 \end{eqnarray}
 \begin{eqnarray}
&\pm&\frac{1}{\lambda_{\Sigma_Q}\lambda_{\Sigma_Q^*}
}\exp{\frac{M_{\Sigma_Q}^2+M_{\Sigma_Q^*}^2-2m_Q^2-2u_0\bar{u}_0M_\rho^2}{2M^2}}\nonumber \\
&&\left\{\frac{8u_0\langle\bar{q}q\rangle f_\rho^{\perp}M_\rho^2
\widetilde{\widetilde{B}}_{\perp}(\bar{u}_0) }{3M^2} -
\frac{2u_0\langle \bar{q}g_s \sigma Gq\rangle
f_\rho^{\perp}M_\rho^2\widetilde{\widetilde{B}}_{\perp}(\bar{u}_0)}{3M^4}\left(1+\frac{m_Q^2}{M^2}\right)
\right\} \, ,
 \end{eqnarray}

\begin{eqnarray}
\rm{G}3_{\Sigma_Q^*\Sigma_Q\rho_0}&=&\pm\frac{1}{\lambda_{\Sigma_Q}\lambda_{\Sigma_Q^*}}\exp{\frac{M_{\Sigma_Q}^2
+M_{\Sigma_Q^*}^2-2u_0\bar{u}_0M_\rho^2}{2M^2}}\nonumber \\
&&\left\{\frac{f_\rho M_\rho
 \left[\widetilde{\phi}_{\parallel}(\bar{u}_0)-\widetilde{g}_{\perp}^{(v)}(\bar{u}_0)\right]}{2\pi^2}M^4E_1(x)\int_0^1 dt t(1-t) e^{-\frac{\widetilde{m}_Q^2}{M^2}}\right.\nonumber\\
 &&- \frac{m_Q^2f_\rho M_\rho
\left[\widetilde{\phi}_{\parallel}(\bar{u}_0)-\widetilde{g}_{\perp}^{(v)}(\bar{u}_0)\right]}{36M^2}\langle\frac{\alpha_sGG}{\pi}\rangle\int_0^1 dt \frac{1-t}{t^2} e^{-\frac{\widetilde{m}_Q^2}{M^2}}\nonumber\\
&& -\frac{f_\rho M_\rho^3  \widetilde{A}(\bar{u}_0)}{8\pi^2}
M^2E_0(x)\int_0^1 dt t
e^{-\frac{\widetilde{m}_Q^2}{M^2}}\nonumber\\
 &&+\frac{m_Q^2f_\rho M_\rho^3 \widetilde{A}(\bar{u}_0)}{144M^4}
\langle \frac{\alpha_sGG}{\pi}
\rangle\int_0^1 dt \frac{1}{t^2} e^{-\frac{\widetilde{m}_Q^2}{M^2}} \nonumber\\
&&+\frac{\widetilde{f}_\rho M_\rho
}{8\pi^2}M^4E_1(x) g_\perp^{(a)}(\bar{u}_0)\int_0^1 dt t(1+t) e^{-\frac{\widetilde{m}_Q^2}{M^2}}\nonumber\\
 &&-\frac{m_Q^2\widetilde{f}_\rho M_\rho
}{144M^2}\langle \frac{\alpha_sGG}{\pi}\rangle g_\perp^{(a)}(\bar{u}_0)\int_0^1 dt \frac{1+t}{t^2} e^{-\frac{\widetilde{m}_Q^2}{M^2}}\nonumber\\
&&  \left.-\frac{u_0f_\rho M_\rho^3 }{2\pi^2}M^2E_0(x) \int_0^1 dt
t\int_0^{u_0} d\alpha_{\bar{u}}
\int_{u_0-\alpha_{\bar{u}}}^{1-\alpha_{\bar{u}}} d\alpha_g
v\frac{\mathcal{A}(\alpha_i)-\mathcal{V}(\alpha_i)} {\alpha_g}
e^{-\frac{\widetilde{m}_Q^2}{M^2}}    \right\} \nonumber
 \end{eqnarray}
\begin{eqnarray}
&\pm&\frac{1}{\lambda_{\Sigma_Q}\lambda_{\Sigma_Q^*}
}\exp{\frac{M_{\Sigma_Q}^2+M_{\Sigma_Q^*}^2-2m_Q^2-2u_0\bar{u}_0M_\rho^2}{2M^2}}\nonumber \\
&&\left\{-\frac{2\langle\bar{q}q\rangle f_\rho^{\perp}
\phi_{\perp}(\bar{u}_0)}{3}M^2E_0(x) + \frac{\langle \bar{q}g_s
\sigma
Gq\rangle f_\rho^{\perp}\phi_{\perp}(\bar{u}_0)}{6}\left(1+\frac{m_Q^2}{M^2}\right) \right.\nonumber\\
 &&- \frac{m_Q^4 f_\rho^{\perp} M_\rho^2 \langle \bar{q}g_s\sigma G q\rangle
A_{\perp}(\bar{u}_0)}{24M^6}  + \frac{\langle\bar{q}q\rangle
f_\rho^{\perp} M_\rho^2 A_{\perp}(\bar{u}_0)}{6} \left(1+\frac{m_Q^2}{M^2}\right)   \nonumber\\
&&\left. +\frac{4\langle\bar{q}q\rangle f_\rho^{\perp}M_\rho^2
\widetilde{\widetilde{B}}_{\perp}(\bar{u}_0) }{3} - \frac{\langle
\bar{q}g_s \sigma Gq\rangle
f_\rho^{\perp}M_\rho^2\widetilde{\widetilde{B}}_{\perp}(\bar{u}_0)}{3M^2}\left(1+\frac{m_Q^2}{M^2}\right)\right\}\,,
 \end{eqnarray}
 where $\bar{u}_0=1-u_0$, $\widetilde{f}_\phi=f_\phi-f_\phi^\perp
\frac{2m_s}{M_\phi}$, $\widetilde{f}_\phi^\perp=f_\phi^\perp-f_\phi
\frac{2m_s}{M_\phi}$, $\widetilde{f}_\rho=f_\rho-f_\rho^\perp
\frac{m_u+m_d}{M_\rho}$,
$\widetilde{f}_\rho^\perp=f_\rho^\perp-f_\rho
\frac{m_u+m_d}{M_\rho}$,  $M_1^2=M_2^2=2M^2$ and
$u_0=\frac{M_1^2}{M_1^2+M_2^2}=\frac{1}{2}$ as
$\frac{M_{B_Q^*}^2}{M_{B_Q^*}^2+M_{B_Q}^2}\approx \frac{1}{2}$,
$v=\frac{u_0-\alpha_{\bar{s}}}{\alpha_g}$
($\frac{u_0-\alpha_{\bar{u}}}{\alpha_g}$),
$\widetilde{m}_Q^2=\frac{m_Q^2}{t}$,
$E_n(x)=1-(1+x+\frac{x^2}{2!}+\cdots+\frac{x^n}{n!})e^{-x}$,
$x=\frac{s_0}{M^2}$;
$\widetilde{\widetilde{f}}(\bar{u}_0)=\int_0^{u_0}du\int_0^u dt
f(1-t)$, $\widetilde{f}(\bar{u}_0)=\int_0^{u_0}du f(1-u)$, the
$f(u)$ denote the light-cone distribution amplitudes, the lengthy
expressions of the light-cone distribution amplitudes
$\phi_{\parallel}(u)$, $\phi_{\perp}(u)$, $A(u)$, $A_\perp(u)$,
$g_\perp^{(v)}(u)$,
  $g_\perp^{(a)}(u)$, $h_{\parallel}^{(s)}(u)$,
  $h_{\parallel}^{(t)}(u)$, $h_3(u)$, $g_3(u)$, $\mathcal{A}(\alpha_i)$, $\mathcal {S}(\alpha_i)$,
  $\widetilde{\mathcal{S}}(\alpha_i)$,
  $\mathcal{T}(\alpha_i)$, $\mathcal{V}(\alpha_i)$ can be found in
  Refs.\cite{VMLC2003,VMLC2007},
  $C(u)=g_3(u)+\phi_{\parallel}(u)-2g_\perp^{(v)}(u)
  $, $B_\perp(u)=h_{\parallel}^{(t)}(u)-\frac{1}{2}\phi_{\perp}(u)-\frac{1}{2}h_3(u)
  $, $C_\perp(u)=h_3(u)-\phi_{\perp}(u) $;
 the denotation $\pm$ correspond to the vertexes
$\Xi_c^{*+}{\Xi'}_c^{+}\rho_0$, $\Xi_b^{*0}{\Xi'}_b^{0}\rho_0$,
$\Sigma_c^{*++}\Sigma_c^{++}\rho_0$,
$\Sigma_b^{*+}\Sigma_b^{+}\rho_0$ and
$\Xi_c^{*0}{\Xi'}_c^{0}\rho_0$, $\Xi_b^{*-}{\Xi'}_b^{-}\rho_0$,
$\Sigma_c^{*0}\Sigma_c^{0}\rho_0$, $\Sigma_b^{*-}\Sigma_b^{-}\rho_0$
respectively. The strong coupling constants $g_1$, $g_2$ and G3 in
the vertexes  $\Sigma_c^{*+}\Sigma_c^{+}\rho_0$ and
$\Sigma_b^{*0}\Sigma_b^{0}\rho_0$ vanish in the isospin symmetry
limit. For some technical details involving the three particle
vector-mesons ($\phi$ and $\rho_0$) light-cone distribution
amplitudes, one can consult Ref.\cite{WangJPG}.

The quark constituents of the vector mesons $\rho_0$ and $\omega$
are $\frac{|\bar{u}u\rangle-|\bar{d}d\rangle}{\sqrt{2}} $ and
$\frac{|\bar{u}u\rangle+|\bar{d}d\rangle}{\sqrt{2}}$ respectively.
For example, the correlation functions
$\Pi_\mu^{\Xi_c^{+*}{\Xi'}_c^{+}\rho/\omega}(p,q)$ and
$\Pi_\mu^{\Xi_c^{0*}{\Xi'}_c^{0}\rho/\omega}(p,q)$ can be decomposed
as
\begin{eqnarray}
\Pi_\mu^{\Xi_c^{+*}{\Xi'}_c^{+}\rho/\omega}(p,q)&=&\frac{i}{\sqrt{2}}\epsilon^{ijk}\epsilon^{i'j'k'}
\int d^4x e^{-i p \cdot x} \gamma_5\gamma^\alpha S_c^{kk'}(-x)
\nonumber \\
 &&Tr\left[ \gamma_\alpha S_{jj'}(-x)\gamma_\mu C\langle 0|u_i(0)\bar{u}_{i'}(x) |\bar{u}u(q)\rangle^T
 C\right] \, ,\nonumber \\
 \Pi_\mu^{\Xi_c^{0*}{\Xi'}_c^{0}\rho/\omega}(p,q)&=&\pm\frac{i}{\sqrt{2}}\epsilon^{ijk}\epsilon^{i'j'k'}
\int d^4x e^{-i p
\cdot x} \gamma_5\gamma^\alpha S_c^{kk'}(-x)\nonumber \\
 &&Tr\left[ \gamma_\alpha S_{jj'}(-x)\gamma_\mu C\langle 0|d_i(0)\bar{d}_{i'}(x) |\bar{d}d(q)\rangle^T
 C\right] \, ,
  \end{eqnarray}
 respectively, where the couplings
  \begin{eqnarray}
\langle 0|u_i(0)\bar{u}_{i'}(0)
|\bar{u}u(q)\rangle\mid_{\rho}&=&\langle 0|d_i(0)\bar{d}_{i'}(0)
|\bar{d}d(q)\rangle\mid_{\rho}\propto f_\rho (f^\perp_\rho)
  M_\rho^n \, , \nonumber \\
  \langle 0|u_i(0)\bar{u}_{i'}(0)
|\bar{u}u(q)\rangle\mid_{\omega}&=&\langle 0|d_i(0)\bar{d}_{i'}(0)
|\bar{d}d(q)\rangle\mid_{\omega}\propto f_\omega (f^\perp_\omega)
  M_\omega^n \, ,
  \end{eqnarray}
  the $n$ is an integer, and the $\pm$ correspond to the vector mesons $\rho_0$ and $\omega$ respectively.
The isospin triplet meson $\rho_0$ and isospin singlet meson
$\omega$ have approximately   degenerate masses, i.e.
$\frac{M_\rho}{ M_\omega}\approx 98.5\%$. The $\omega$-meson
light-cone distribution amplitudes have not been explored yet, we
assume that the vector mesons $\rho_0$ and $\omega$ have similar
light-cone distribution amplitudes, and take the approximation
$M_\omega=M_\rho$, $f_\omega=f_\rho$, $f^\perp_\omega=f^\perp_\rho$,
$\cdots$ for the hadronic parameters and obtain the strong coupling
constants involving the vector meson $\omega$ by symmetry
considerations, which are shown in Table 3. Such an approximation is
not crude, for example, if we study the masses and decay constants
of the vector mesons $\rho_0$ and $\omega$ using the interpolating
currents $J^\rho_\mu=\frac{1}{\sqrt{2}}(\bar{u}\gamma_\mu
u-\bar{d}\gamma_\mu d)$ and
$J^\omega_\mu=\frac{1}{\sqrt{2}}(\bar{u}\gamma_\mu
u+\bar{d}\gamma_\mu d)$ respectively with the QCD sum rules, the
resulting values are almost degenerate.

\subsection*{Appendix\,\,B}
The spectral densities of the heavy baryon states $\Xi_Q^*$,
$\Xi'_Q$, $\Sigma_Q^*$ and $\Sigma_Q$ at the level of quark-gluon
degrees of freedom,
\begin{eqnarray}
\rho^A_{\Xi_Q^*}(s)&=&\frac{1}{128\pi^4}\int_{t_i}^1dt
t(t+2)(1-t)^2(s-\widetilde{m}_Q^2)^2+\frac{m_s\langle\bar{s}s\rangle}{16\pi^2}\int_{t_i}^1
dt t^2 -\frac{m_s\langle\bar{q}q\rangle}{8\pi^2}\int_{t_i}^1 dt t
\nonumber\\
&&+\frac{m_s\langle\bar{s}g_s\sigma Gs\rangle}{96\pi^2}\int_0^1dt
t\delta
(s-\widetilde{m}_Q^2)+\frac{m_s\left[3\langle\bar{q}g_s\sigma
Gq\rangle-\langle\bar{s}g_s\sigma
Gs\rangle\right]}{96\pi^2}\delta(s-m_Q^2) \nonumber\\
&&-\frac{m_Q^2}{1152\pi^2}\langle \frac{\alpha_sGG}{\pi}\rangle
\int_0^1 dt\frac{(t+2)(1-t)^2}{t^2}\delta
(s-\widetilde{m}_Q^2)+\frac{\langle\bar{q}q\rangle\langle\bar{s}s\rangle}{6}\delta(s-m_Q^2)
\nonumber \\
&&-\frac{1}{384\pi^2}\langle \frac{\alpha_sGG}{\pi}\rangle
\int_{t_i}^1 dt t(2-t) \, ,
\end{eqnarray}

\begin{eqnarray}
\rho^B_{\Xi_Q^*}(s)&=&\frac{m_Q}{128\pi^4}\int_{t_i}^1dt
(t+2)(1-t)^2(s-\widetilde{m}_Q^2)^2+\frac{m_sm_Q\langle\bar{s}s\rangle}{16\pi^2}\int_{t_i}^1
dt t -\frac{m_sm_Q\langle\bar{q}q\rangle}{8\pi^2}\int_{t_i}^1 dt
\nonumber\\
&&+\frac{m_sm_Q\langle\bar{s}g_s\sigma Gs\rangle}{96\pi^2}\int_0^1dt
\delta
(s-\widetilde{m}_Q^2)+\frac{m_sm_Q\left[3\langle\bar{q}g_s\sigma
Gq\rangle-\langle\bar{s}g_s\sigma
Gs\rangle\right]}{96\pi^2}\delta(s-m_Q^2) \nonumber\\
&&-\frac{m_Q}{1152\pi^2}\langle \frac{\alpha_sGG}{\pi}\rangle
\int_{t_i}^1
dt\frac{3t^4-8t^3+3t^2+9t-4}{t^2}+\frac{m_Q\langle\bar{q}q\rangle\langle\bar{s}s\rangle}{6}\delta(s-m_Q^2)
\nonumber \\
&&-\frac{m_Q}{1152\pi^2}\langle \frac{\alpha_sGG}{\pi}\rangle
\int_0^1 dt\frac{t^3-3t+2}{t}\widetilde{m}_Q^2\delta
(s-\widetilde{m}_Q^2) \, ,
\end{eqnarray}

\begin{eqnarray}
\rho^A_{\Sigma_Q^*}(s)&=&\frac{1}{128\pi^4}\int_{t_i}^1dt
t(t+2)(1-t)^2(s-\widetilde{m}_Q^2)^2+\frac{\langle\bar{q}q\rangle^2}{6}\delta(s-m_Q^2)
\nonumber\\
&&-\frac{m_Q^2}{1152\pi^2}\langle \frac{\alpha_sGG}{\pi}\rangle
\int_0^1 dt\frac{(t+2)(1-t)^2}{t^2}\delta (s-\widetilde{m}_Q^2)
\nonumber \\
&&-\frac{1}{384\pi^2}\langle \frac{\alpha_sGG}{\pi}\rangle
\int_{t_i}^1 dt t(2-t) \, ,
\end{eqnarray}

\begin{eqnarray}
\rho^B_{\Sigma_Q^*}(s)&=&\frac{m_Q}{128\pi^4}\int_{t_i}^1dt
(t+2)(1-t)^2(s-\widetilde{m}_Q^2)^2+\frac{m_Q\langle\bar{q}q\rangle^2}{6}\delta(s-m_Q^2)
\nonumber\\
&&-\frac{m_Q}{1152\pi^2}\langle \frac{\alpha_sGG}{\pi}\rangle
\int_{t_i}^1 dt\frac{3t^4-8t^3+3t^2+9t-4}{t^2}
\nonumber \\
&&-\frac{m_Q}{1152\pi^2}\langle \frac{\alpha_sGG}{\pi}\rangle
\int_0^1 dt\frac{t^3-3t+2}{t}\widetilde{m}_Q^2\delta
(s-\widetilde{m}_Q^2) \, ,
\end{eqnarray}

\begin{eqnarray}
\rho^A_{\Xi'_Q}(s)&=&\frac{1}{32\pi^4}\int_{t_i}^1dt
t(1-t)^3(s-\widetilde{m}_Q^2)(5s-3\widetilde{m}_Q^2)-\frac{m_s\langle\bar{q}q\rangle}{4\pi^2}\int_{t_i}^1
dt t\nonumber\\
&&+\frac{m_s\langle\bar{s}s\rangle}{4\pi^2}\int_{t_i}^1 dt
t(1-t)\left[3+s\delta (s-\widetilde{m}_Q^2)\right]
\nonumber\\
&&-\frac{m_s\langle\bar{s}g_s\sigma Gs\rangle}{24\pi^2}\int_0^1dt
t\left[2+\frac{s}{M^2}\right]\delta
(s-\widetilde{m}_Q^2) \nonumber\\
&&+\frac{m_s\langle\bar{q}g_s\sigma Gq\rangle}{16\pi^2}\delta
(s-m_Q^2)+\frac{\langle\bar{q}q\rangle\langle\bar{s}s\rangle}{3}\delta(s-m_Q^2)\nonumber \\
&& +\frac{1}{96\pi^2}\langle \frac{\alpha_sGG}{\pi}\rangle
\int_{t_i}^1 dt (4-5t)+\frac{1}{96\pi^2}\langle
\frac{\alpha_sGG}{\pi}\rangle \int_{t_i}^1 dt
(1-t)\widetilde{m}_Q^2\delta
(s-\widetilde{m}_Q^2)\nonumber \\
&& -\frac{m_Q^2}{288\pi^2}\langle \frac{\alpha_sGG}{\pi}\rangle
\int_0^1 dt\frac{(1-t)^3}{t^2}\left[2+\frac{s}{M^2}\right]\delta
(s-\widetilde{m}_Q^2) \, ,
\end{eqnarray}

\begin{eqnarray}
\rho^B_{\Xi'_Q}(s)&=&\frac{3m_Q}{64\pi^4}\int_{t_i}^1dt
(1-t)^2(s-\widetilde{m}_Q^2)^2-\frac{m_sm_Q\langle\bar{q}q\rangle}{2\pi^2}\int_{t_i}^1
dt+\frac{m_sm_Q\langle\bar{s}s\rangle}{8\pi^2}\int_{t_i}^1 dt
\nonumber\\
&&+\frac{m_sm_Q\left[6\langle\bar{q}g_s\sigma
Gq\rangle-\langle\bar{s}g_s\sigma Gs\rangle \right]}{48\pi^2}\delta
(s-m_Q^2) +\frac{2m_Q\langle\bar{q}q\rangle\langle\bar{s}s\rangle}{3}\delta(s-m_Q^2)\nonumber\\
&&+\frac{m_Q}{192\pi^2}\langle \frac{\alpha_sGG}{\pi}\rangle
\int_{t_i}^1 dt \left[ -3-2t+\frac{2}{t^2}\right]\nonumber \\
&& -\frac{m_Q}{192\pi^2}\langle \frac{\alpha_sGG}{\pi}\rangle
\int_0^1 dt\frac{(1-t)^2}{t}\widetilde{m}_Q^2\delta
(s-\widetilde{m}_Q^2) \, ,
\end{eqnarray}

\begin{eqnarray}
\rho^A_{\Sigma_Q}(s)&=&\frac{1}{32\pi^4}\int_{t_i}^1dt
t(1-t)^3(s-\widetilde{m}_Q^2)(5s-3\widetilde{m}_Q^2)+\frac{\langle\bar{q}q\rangle^2}{3}\delta(s-m_Q^2)
\nonumber \\
&& +\frac{1}{96\pi^2}\langle \frac{\alpha_sGG}{\pi}\rangle
\int_{t_i}^1 dt (4-5t)+\frac{1}{96\pi^2}\langle
\frac{\alpha_sGG}{\pi}\rangle \int_{t_i}^1 dt
(1-t)\widetilde{m}_Q^2\delta
(s-\widetilde{m}_Q^2)\nonumber \\
&& -\frac{m_Q^2}{288\pi^2}\langle \frac{\alpha_sGG}{\pi}\rangle
\int_0^1 dt\frac{(1-t)^3}{t^2}\left[2+\frac{s}{M^2}\right]\delta
(s-\widetilde{m}_Q^2) \, ,
\end{eqnarray}

\begin{eqnarray}
\rho^B_{\Sigma_Q}(s)&=&\frac{3m_Q}{64\pi^4}\int_{t_i}^1dt
(1-t)^2(s-\widetilde{m}_Q^2)^2
+\frac{2m_Q\langle\bar{q}q\rangle^2}{3}\delta(s-m_Q^2)\nonumber\\
&&+\frac{m_Q}{192\pi^2}\langle \frac{\alpha_sGG}{\pi}\rangle
\int_{t_i}^1 dt \left[ -3-2t+\frac{2}{t^2}\right]\nonumber \\
&& -\frac{m_Q}{192\pi^2}\langle \frac{\alpha_sGG}{\pi}\rangle
\int_0^1 dt\frac{(1-t)^2}{t}\widetilde{m}_Q^2\delta
(s-\widetilde{m}_Q^2) \, ,
\end{eqnarray}
where $\widetilde{m}_Q^2=\frac{m_Q^2}{t}$, $t_i=\frac{m_Q^2}{s}$.


\begin{thebibliography}{99}
\bibitem{ReviewH1}  J. G. Koerner, D. Pirjol and M. Kraemer, Prog. Part. Nucl. Phys. {\bf 33} (1994)
787.

\bibitem{ReviewH2} F. Hussain, G. Thompson and J. G. Koerner, hep-ph/9311309.

\bibitem{PDG} C. Amsler et al,  Phys. Lett. {\bf B667} (2008)  1.

\bibitem{ShortRV1} T. Lesiak, hep-ex/0612042.

\bibitem{ShortRV2} J. L. Rosner, J. Phys. Conf. Ser. {\bf 69} (2007) 012002.

\bibitem{ShortRV3}  M. Paulini, arXiv:0906.0808.

\bibitem{WangVMD} Z. G. Wang,  Phys. Rev. {\bf D75} (2007) 034013.


\bibitem{Wang0809Omega} Z. G. Wang,  Eur. Phys. J. {\bf C61} (2009)
321.

\bibitem{Wang0704} Z. G. Wang, Eur. Phys. J. {\bf C54} (2008) 231.


\bibitem{Wang0909} Z. G. Wang, arXiv:0909.4144.

\bibitem{Aliev0901} T. M. Aliev, K. Azizi and A. Ozpineci, Phys. Rev. {\bf D79} (2009) 056005.

\bibitem{LCSR89}
I. I. Balitsky, V. M. Braun and A. V. Kolesnichenko, Nucl. Phys.
{\bf B312} (1989) 509.

\bibitem{LCSR} V. L. Chernyak and A. R. Zhitnitsky, Phys. Rept. {\bf 112} (1984)
173.

\bibitem{LCSRreview} P. Colangelo and A. Khodjamirian, hep-ph/0010175.


\bibitem{SVZ79} M. A. Shifman, A. I. Vainshtein and V. I. Zakharov,
Nucl. Phys. {\bf B147} (1979) 385, 448.

\bibitem{PRT85} L. J. Reinders, H. Rubinstein and S. Yazaki, Phys. Rept. {\bf
127} (1985) 1.

\bibitem{Zhu99} S. L. Zhu, Phys. Rev. {\bf C59} (1999) 435.


\bibitem{Wang0701} Z. G. Wang,  Phys. Rev. {\bf D75} (2007) 054020.


\bibitem{Aliev0905}  T. M. Aliev, A. Ozpineci, M. Savci and V. S. Zamiralov, Phys. Rev. {\bf D80} (2009) 016010.


\bibitem{Aliev2006} T. M. Aliev, A. Ozpineci,  S. B. Yakovlev and V. Zamiralov,  Phys. Rev. {\bf D74} (2006) 116001.

\bibitem{Aliev0908} T. M. Aliev, K. Azizi, A. Ozpineci and M. Savci,
 Phys. Rev. {\bf D80} (2009) 096003.


\bibitem{Wang0707} Z. G. Wang,  Eur. Phys. J. {\bf C57} (2008) 711.

\bibitem{Wang0809} Z. G. Wang, Eur. Phys. J. {\bf C61} (2009) 299.


\bibitem{Erkol2006} G. Erkol, R. G. E. Timmermans and T. A. Rijken,  Phys. Rev. {\bf C74} (2006) 045201.

\bibitem{Zhu0909} P. Z. Huang, H. X Chen and  S. L. Zhu,
arXiv:0909.5551.

\bibitem{ANN1973} H. F. Jones and M. D. Scadron,  Ann. Phys. {\bf 81} (1973) 1.


\bibitem{VMLC981} P. Ball,  V. M. Braun, Y. Koike and K. Tanaka,  Nucl. Phys. {\bf B529} (1998) 323.

\bibitem{VMLC982} P. Ball and V. M. Braun,  Nucl. Phys. {\bf B543} (1999) 201.

\bibitem{VMLC2003}  P. Ball and M. Boglione,  Phys. Rev. {\bf D68} (2003) 094006.

\bibitem{VMLC2007} P. Ball,  V. M. Braun and A. Lenz, JHEP {\bf 0708} (2007) 090.


\bibitem{Ioffe2005} B. L. Ioffe, Prog. Part. Nucl. Phys. {\bf 56} (2006)
232.

\bibitem{Huang08} J. R. Zhang and M. Q. Huang, Phys. Rev. {\bf D78} (2008) 094015.

\bibitem{Narison09} M. Albuquerque,  S. Narison and M. Nielsen,
arXiv:0904.3717.

\bibitem{Jaffe2003} R. L. Jaffe and  F. Wilczek, Phys. Rev. Lett. {\bf 91} (2003) 232003.


\bibitem{Jaffe2004} R. L. Jaffe, Phys. Rept. {\bf 409} (2005) 1.


\bibitem{WangJPG} Z. G. Wang, J. Phys. {\bf G34} (2007) 753.


\bibitem{HHCPT-1} M. C. Banuls, A. Pich and I. Scimemi, Phys. Rev. {\bf D61} (2000)
094009.

\bibitem{HHCPT-2} H. Y. Cheng, C. Y. Cheung, G. L. Lin, Y. C. Lin, T. M. Yan and H. L. Yu,
Phys. Rev. {\bf D47} (1993) 1030.

\bibitem{Heavy-Light} S. Tawfiq, J. G. Koerner and P. J. O'Donnell, Phys. Rev. {\bf D63} (2001)
034005.

\bibitem{R3QM-1}  M. A. Ivanov, J. G. Korner and V. E. Lyubovitskij, Phys.
Lett. {\bf  B448}  (1999) 143.

\bibitem{R3QM-2} M. A. Ivanov, J. G. Korner and V. E. Lyubovitskij, Phys. Rev. {\bf D60} (1999) 094002.


\bibitem{CQMwidth} A. Majethiya, B. Patel and P. C. Vinodkumar, Eur. Phys. J. {\bf A42} (2009)
213.

\bibitem{Dey1994} J. Dey, M. Dey, V. Shevchenko and P. Volkovitsky, Phys. Lett. {\bf B337} (1994) 185.

\bibitem{BESIII} D. M. Asner et al, arXiv:0809.1869.

\bibitem{PANDA} M. F. M. Lutz et al, arXiv:0903.3905.

\bibitem{LHC}  G. Kane and A. Pierce, "Perspectives On LHC Physics",
World Scientific Publishing Company,  2008.
\end{thebibliography}
\end{document}